\documentclass[aps,reprint,superscriptaddress,prb]{revtex4-2}
\usepackage[utf8]{inputenc}
\usepackage[T1]{fontenc}
\usepackage{amsmath,amssymb,graphicx,color}
\graphicspath{{figures/}}
\usepackage[bookmarks=false]{hyperref}
\usepackage[usenames, dvipsnames]{xcolor}
\usepackage{subcaption}
\usepackage[super]{nth}
\usepackage[normalem]{ulem}
\def\be{\begin{equation}}
\def\ee{\end{equation}}
\def\ba{\begin{eqnarray}}
\def\ea{\end{eqnarray}}
\def\CVS{CsV$_3$Sb$_5$~}
\def\KVS{KV$_3$Sb$_5$~}
\def\RVS{RbV$_3$Sb$_5$~}
\def\AVS{AV$_3$Sb$_5$~}
\def\SVS{ScV$_6$Sn$_6$~}

\def\muSR{$\mu$SR~}

\begin{document}

\title{ High Resolution Polar Kerr Effect Studies of CsV$_3$Sb$_5$ and ScV$_6$Sn$_6$ Below the Charge Order Transition}

\author{David R. Saykin}
\affiliation{Geballe Laboratory for Advanced Materials, Stanford University, Stanford, CA 94305, USA.}
\affiliation{Stanford Institute for Materials and Energy Sciences, SLAC National Accelerator Laboratory, 2575 Sand Hill Road, Menlo Park, CA 94025, USA.}
\affiliation{Department of Physics, Stanford University, Stanford, CA 94305, USA.}

\author{Qianni Jiang}
\affiliation{Stanford Institute for Materials and Energy Sciences, SLAC National Accelerator Laboratory, 2575 Sand Hill Road, Menlo Park, CA 94025, USA.}
\affiliation{Department of Applied Physics, Stanford University, Stanford, CA 94305, USA.}

\author{Zhaoyu Liu}
\affiliation{Department of Physics, University of Washington, Seattle, Washington, 98195, USA}

\author{Chandra Shekhar}
\affiliation{Max Planck Institute for Chemical Physics of Solids, 01187 Dresden, Germany.}

\author{Claudia Felser}
\affiliation{Max Planck Institute for Chemical Physics of Solids, 01187 Dresden, Germany.}

\author{Jiun-Haw Chu}
\affiliation{Department of Physics, University of Washington, Seattle, Washington, 98195, USA}

\author{Aharon Kapitulnik}
\email{aharonk@stanford.edu}
\affiliation{Geballe Laboratory for Advanced Materials, Stanford University, Stanford, CA 94305, USA.}
\affiliation{Stanford Institute for Materials and Energy Sciences, SLAC National Accelerator Laboratory, 2575 Sand Hill Road, Menlo Park, CA 94025, USA.}
\affiliation{Department of Applied Physics, Stanford University, Stanford, CA 94305, USA.}
\affiliation{Department of Physics, Stanford University, Stanford, CA 94305, USA.}


\begin{abstract}
We report high resolution polar Kerr effect measurements on \CVS and \SVS single crystals in search for signatures of spontaneous polar Kerr effect (PKE) below the charge order transitions of these materials.  Utilizing two separate zero-area loop Sagnac interferometers operating at 1550 nm and 830 nm wavelengths, we studied the temperature dependence of possible PKE after training with magnetic field. While a finite field Kerr measurement yielded optical rotation expected from the Pauli susceptibility of the itinerant carriers, no signal was detected at zero-field  to within the noise floor limit of the apparatus of below $\sim$100 nanoradians. Simultaneous  coherent reflection measurements  confirm the sharpness of the charge order transition in the same optical volume as the Kerr measurements.  Application of strain to reveal a hidden flux-ordered magnetic state did not result in a finite Kerr effect. 
\end{abstract}
\pacs{NaN}

\maketitle

\section{Introduction}

Kagom\'e lattice-based quantum materials feature corner-sharing triangles arranged in a hexagonal cell that lead to a wide range of unique electronic phases controlled by correlations and topology (see e.g. \cite{Wilson2024,Wang2025,Li2025}). In particular, emerging flat bands and van Hove singularities (vHSs) promote strong electron-electron interactions and electron-phonon coupling that may result in structural instabilities as well as competing electronic orders such as charge density waves (CDWs), superconductivity, and nematicity.  A particularly interesting pair of compounds to compare are  \CVS where a charge-density wave (CDW) state emerges below approximately $T_\text{CDW} = 94$ K, characterized by a lattice reconstruction within the Vanadium plane \cite{Ortiz2020},  \SVS exhibits a CDW state below $T_\text{CDW} = 92$ K, associated primarily with modulation of Sc and Sn atom along $c$--axis \cite{Arachchige2022}. While \CVS exhibits superconductivity at low temperatures ($T_c\lesssim 4$ K), \SVS remains a normal metal with moderate residual resistivity of $\sim 20~\mu\Omega$-cm. 

\begin{figure}[h]
	\includegraphics[width=\columnwidth]{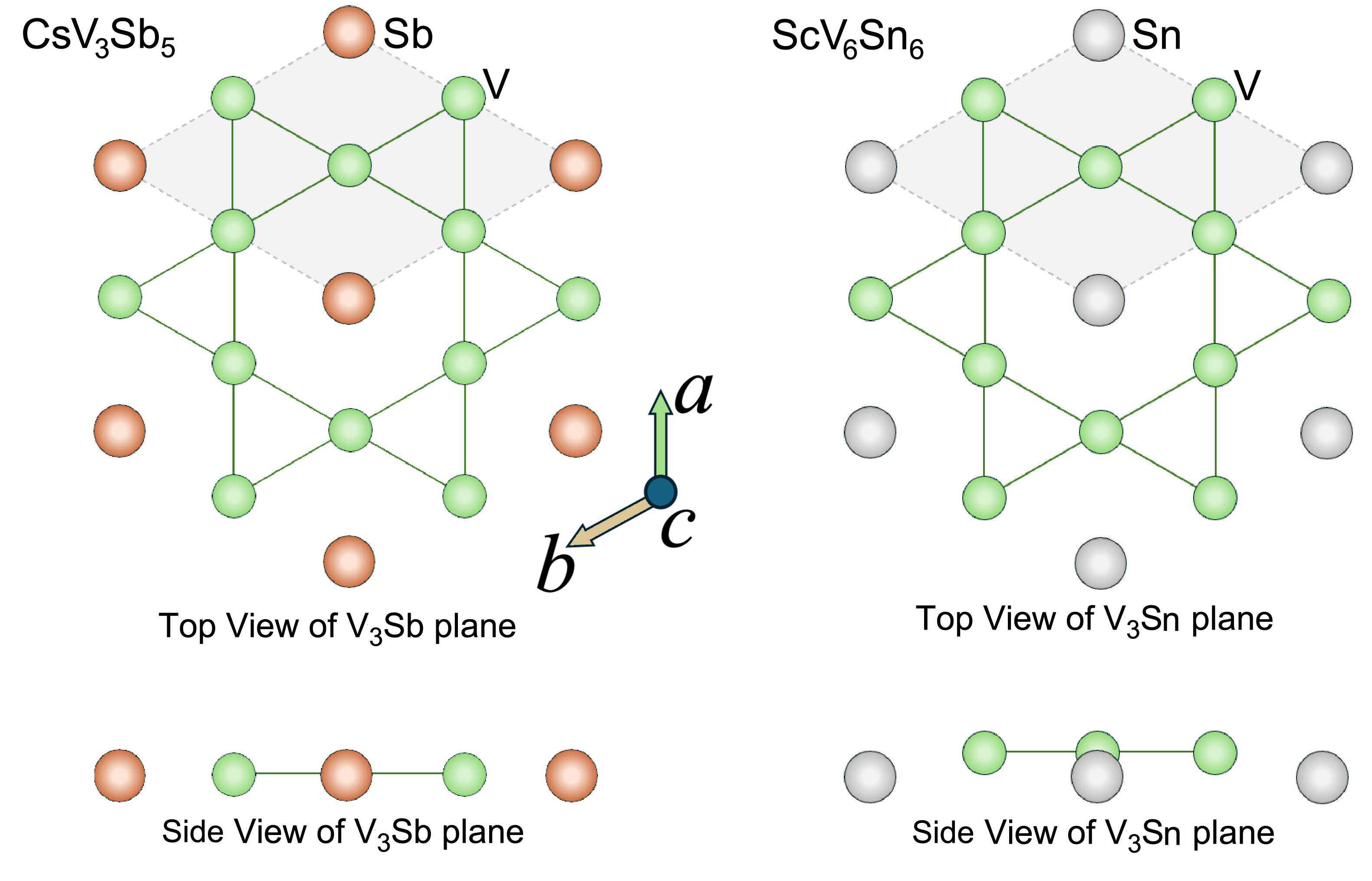}
	\caption{Corner-sharing triangles of vanadium atoms constitute the Kagom\'e lattice planes of CsV$_3$Sb$_5$ (left) and ScV$_6$Sn$_6$ (right). Side views highlight the relative displacement of the vanadium and Sn planes in ScV$_6$Sn$_6$. Shaded area corresponds to the rhombus- shaped unit cell.}
	\label{structure}
\end{figure}

A lingering issue associated with kagom\'e systems exhibiting charge order transitions originates from the theoretical prediction of co-occurrence of time-reversal symmetry breaking (TRSB)  associated with a flux-ordered magnetic state and the CDW transition \cite{Feng2021,Denner2021,Li2024}. Such chiral flux phases represent an intriguing state of matter, potentially hosting exotic transport and optical responses due to nontrivial electronic topology and orbital magnetism. In particular coupling to strain was suggested to unveil TRSB-originated effects \cite{Guo2022,Guo2025,Alkorta2025}, a property that needs further experimental scrutiny.

Experimental evidence regarding time-reversal symmetry breaking (TRSB) in \AVS ($A$ = Cs, Rb, K) metals has been controversial. Initial scanning tunneling microscopy (STM) studies reported chirality switching of CDW peaks under high applied magnetic field, indicative of a TRS-broken chiral state \cite{Wang2021,Jiang2021,Shumiya2021}. However, subsequent STM studies of \CVS with spin-polarized tips found no evidence of chirality switching or local magnetic moments \cite{Li2022CVS} challenging earlier interpretations. Similarly, an independent STM study of \KVS reported the absence of sensitivity of CDW peaks to magnetic field \cite{Li2022KVS}, but at the same time recent report on \RVS has confirmed the possibility of chirality flipping with the sign of magnetic field \cite{Xing2024}.

Muon spin relaxation ($\mu$SR) experiments have also provided conflicting results. Studies of \CVS observed enhanced internal magnetic fields below the CDW transition \cite{Khasanov2022,Shan2022} and interpreted it as evidence of TRSB. Similar \muSR evidence for hidden magnetism was reported in \SVS \cite{Guguchia2023}. At the same time \muSR experiments on \KVS \cite{Kenney2021} attribute observed changes in relaxation rates at the CDW transition to nuclear rather than electronic fields, thus concluding that no evidence of TRSB is present.  Transport measurements have revealed an appearance of non-linear-in-$H$ behavior below the CDW transition in \CVS and related compounds, which was interpreted as a precursor to anomalous Hall effect indicative of spontaneous orbital magnetism \cite{Yu2021,Yang2020,Yi2024,Mozaffari2024}. However, recent detailed studies of \CVS propose an alternative explanation: these apparent anomalies could arise from high-mobility low-density Fermi pockets opening below $T_\textbf{CDW}$ and specific Fermi surface reconstructions rather than intrinsic TRSB phenomena \cite{Liu2025b,Koshelev2024}.

Magneto-optic (MO) experiments have provided some of the most intriguing, yet contentious, results regarding TRSB. Here, early experiments \cite{Wu2021,Xu2022} searching for either magneto-optic polar Kerr effect (PKE), or its associated ellipticity, also known as reflective circular dichroism (RCD) were performed through analysis of the polarization and amplitude of linearly polarized light reflected from \CVS samples. Both experiments judged to yield a very large effect, of order $\sim50~\mu$rad for PKE and $\sim1$ millirad RCD signals, which would deem \CVS similar in its MO response to hard ferromagnets \cite{Oppeneer2001}. Such a result would also be at odd with optical measurements, which did not find any special optical resonance associated with the CDW at the vicinity of $\sim800$ nm that could explain such a strong response  \cite{Uykur2021,Zhou2021}. By contrast, Saykin {\it et al.} \cite{Saykin2023} performed high-resolution Sagnac interferometry measurements at a wavelength of 1550 nm, finding no observable spontaneous Kerr rotation within the $30$ nrad noise floor over the $d = 10$ $\mu$m beam spot size. Furthermore, the center frequency of these measurements fell well within the full-width at half maximum of the most prominent Lorentz peak associated with the CDW transition centered at $\sim1750$ nm, thus expecting a strong response-if existed. Further testing the positive results at 1550 nm demonstrated a large optical rotation response of linearly polarized light, which was largely isotropic and independent of magnetic fields, thus unrelated to TRSB \cite{Farhang2023}. 

In the present paper we complete our search for TRSB through searching for a finite PKE response in \CVS by performing measurements using a newly constructed Zero-Area Sagnac Interferometer (ZALSI) at 830 nm wavelength. We further perform similar measurements at both 1550 nm and 830 nm wavelengths on \SVS, where TRSB was similarly deduced from $\mu$SR  \cite{Guguchia2023} and Hall effect \cite{Yi2024,Mozaffari2024} measurements. Our primary result is that neither \CVS nor \SVS show a discernible PKE signal to within $\sim50$ nanorad. Additionally, uniaxial strain of magnitude
±0.1\% does not induce spontaneous Kerr signals in \CVS, despite significant strain sensitivity previously reported in transport \cite{Guo2022,Guo2025} and STM results on \RVS \cite{Xing2024}, with the assumption that this system is similar in behavior to \CVS. Our findings significantly constrain the potential magnitude of the orbital flux phase and challenge theoretical models that predict large observable TRSB effects, particularly arguing for a large Kerr signal \cite{Liu2025a}.

\section{Methods}

\begin{figure}[h]
	\includegraphics[width=\columnwidth]{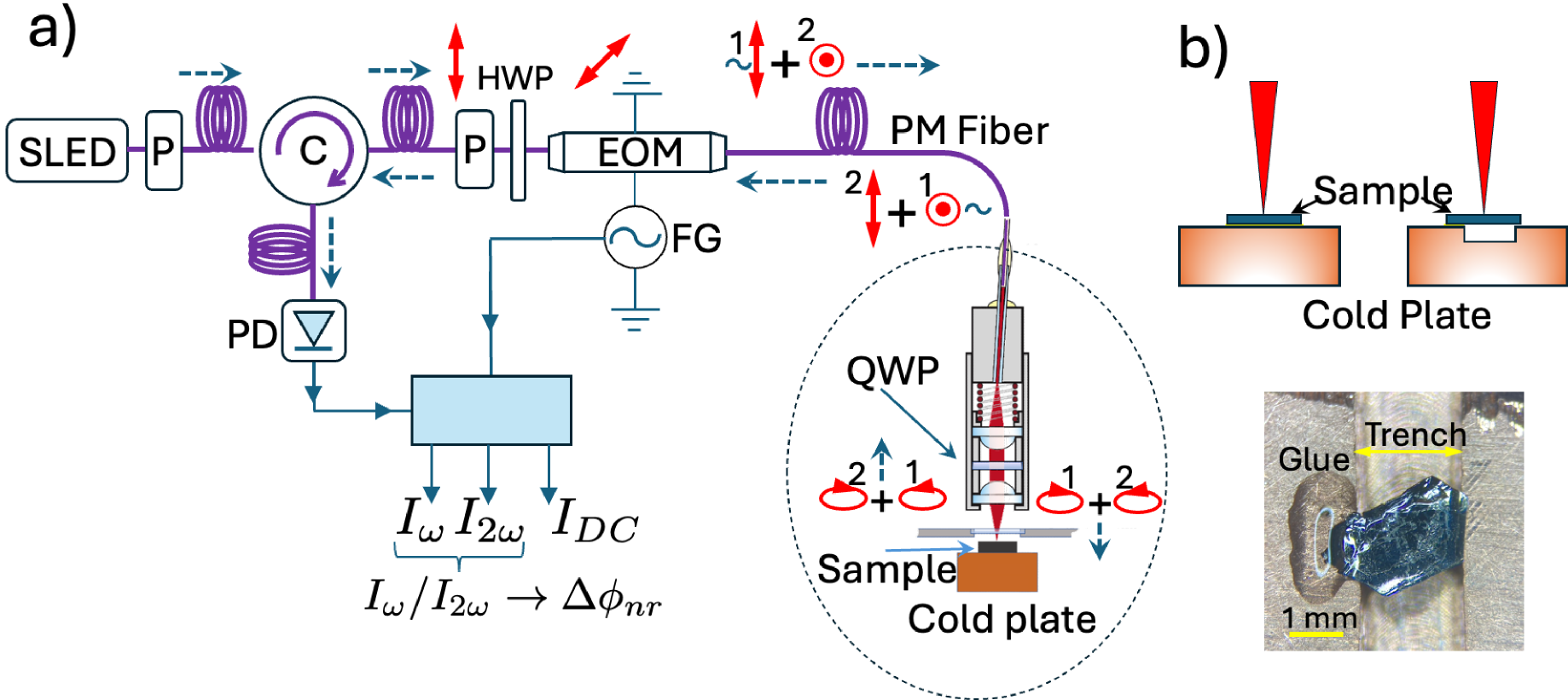}
	\caption{Experimental setup. a) Schematic of the Zero-Area Loop Sagnac Interferometer system. Light emitted from a SLED is polarized (P), then going through a circulator (C) and vertically polarized in (P). Half waveplate rotates the polarization at 45$^\circ$ and only the vertical component is modulated at the electro-optic modulator (EOM) at frequency $\omega$, controlled by a function generator (FG). Upon exiting the EOM, the two, now incoherent components (marked as ``1'' and ``2'', are launched into the two axes of a polarization maintaining (PM). The two beams enter the probe end optics assembly (enlarged within the dashed-lined ellipse), where a quarter waveplate (QWP) transforms the two linear polarizations into right and left circular polarizations. Upon reflection from the sample the two beams exchange circular polarization role. In the presence of birefringent these polarizations become slightly elliptical, and in the presence of TRSB a non reciprocal phase shift, $\Delta\phi_{nr}$ is acquired. Moving back in the same optical path, the two beams coherently combine at the detector and the output signal is analyzed to extract $\theta_K=\Delta\phi_{nr}$.  b)  Top shows two ways of mounting of a \CVS crystal either flat on the cold plate or over a trench, affixed at one side only, while bottom shows a photo of a mounted sample over a trench.}
	\label{methods}
\end{figure}
High-resolution measurement of magneto-optical Kerr response is enabled by utilizing a zero-area-loop Sagnac interferometer (ZALSI), first reported by Xia {\it et al.}  \cite{Xia2006a}. At the heart of the ZALSI (Fig.~\ref{methods}a is a Sagnac loop \cite{Sagnac1913} featuring two counter-propagating beams traveling in the two orthogonal axes of a $\sim10$ m polarization maintaining optical fiber. A quarter waveplate (QWP) at the end of the fiber converts the two beams into right and left circularly polarized light that are focused onto a small interaction region on the sample. The reflected light beams are converted back to linear polarizations with exchanged linear polarization states, thus completing the loop in which both counter propagating beams experience the exact same optical path. Owing to the reciprocity of the apparatus, a non reciprocal phase shift, $\Delta\phi_{nr}$, will appear at the detector only if time reversal symmetry is broken through the interaction of the two circularly polarized beams with the sample. To extract $\Delta\phi_{nr}$, the relative phase of the two counter propagating beams is modulated at a proper frequency $\omega$ (see Supplementary Material \cite{supplement}) such that the signal at 0$\omega$, 1$\omega$, and 2$\omega$ denoted as $I_{DC}$, $I_{\omega}$ and $I_{2\omega}$ can yield the Kerr signal through ${\rm tan}\theta_K=CI_{\omega}/I_{2\omega}$ ($C$ is a known constant). In addition, both, $I_{2\omega}$ and $I_{DC}$ as well as their ratio are a sensitive measure of any linear and/or circular birefringence effects, including reciprocal effects, present in the sample.  For example, it was previously used to detect the LTO to LTT transition in LBCO with exquisite resolution \cite{Karapetyan2012}. For a detailed description of the apparatus see e.g.: Kapitulnik, {\it et al.}\cite{Kapitulnik2009}.  Some of our notable accomplishments include the study of TRSB in Sr$_2$RuO$_4$ \cite{Xia2006b} and UPt$_3$ \cite{Schemm2014}, the discovery of the limit for ferromagnetism in thin SrRuO$_3$ films \cite{Xia2009b}, the inverse proximity effect in ferromagnet/superconductor bilayers \cite{Xia2009a}, and subtle ferromagnetism beyond the dome in LCCO-cuprates \cite{Sarkar2020}. 

High quality single crystals of \CVS were synthesized using the self-flux method. The flux is a eutectic Cs$_x$Sb$_y$  mixed with VSb$_2$ as described in \cite{Ortiz2020}, while  single crystals of \SVS were grown using the Sn-flux method, as previously reported  \cite{Arachchige2022}, yielding hexagon-shaped single crystals.

Previous measurements on \CVS suggested that strain may alter the CDW transition and the subsequent emerging phases \cite{Guo2025}. Thus, samples were mounted onto the cold finger in two different configurations. First directly attached to a flat cold plate and second only to one side of a trench carved in the cold plate (see Fig.~\ref{methods}b. No difference in optical measurements was found in the two mounting configurations.  However, to directly test the effect of strain on the Kerr results, samples were cut in the desired direction and then mounted on a  ``Razorbill'' commercial cryogenic uniaxial strain cell, CS100. Following STM results on \RVS \cite{Xing2024}, we also use a small magnetic field ($\pm25$ mT) to train the samples in the presence of uniaxial tensile or compressional stress in the $a$-direction (See Fig.~\ref{structure}). Measurements were done while warming up the sample in zero-magnetic field at fixed applied strain.

\section{Results and Discussion}

\begin{figure}[h]
	\includegraphics[width=\columnwidth]{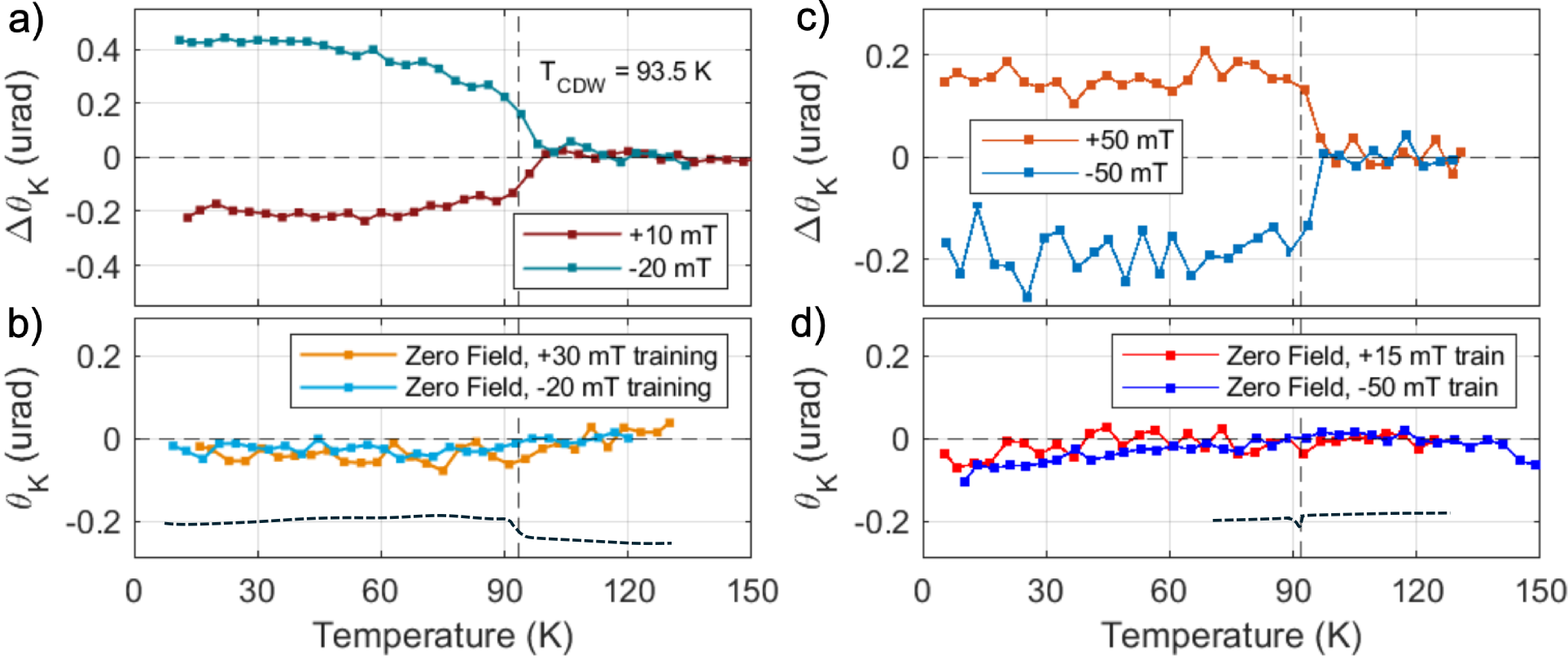}
	\caption{Polar Kerr Effect measurements at 1550 nm wavelength: a,b)  on \CVS, and c,d) on \SVS. Measurements in magnetic field (panels (a and (c)) track the transition with signal of order of the respective Pauli susceptibility \cite{Saykin2023}. Zero-field warmup measurements after training the sample in a field while cooling it through the CDW transition (panels (a and (c)) Show no evidence for a spontaneous Kerr effect (see text). Dashed lines in panels (b) and (d) track the DC component of the reflectivity (in arbitrary units) measured simultaneously with the PKE, which reaffirms that the optical volume that we test indeed undergo the CDW transition.}
	\label{1550}
\end{figure}

Overall,  our ZALSI results on \CVS and \SVS follow the initial study on \CVS \cite{Saykin2023}, where recording $I_{DC}$, $I_{\omega}$ and $I_{2\omega}$ as a function of temperature and magnetic field training effects yield information about possible spontaneous PKE in the samples as well as track the CDW transition through the emergence of any birefringence effects (see methods section).  Figs.~\ref{1550}a,b show data taken using a 1550 nm ZALSI. Similar to ref.~ \cite{Saykin2023}, a finite Kerr response appears in the presence of magnetic field, similar to susceptibility measurements on this material, while at zero applied magnetic field, no PKE signal is observed through the CDW transition within the instrument resolution of $\sim30$ to $50$ nanorad. Figs.~\ref{1550}c,d show similar data, also at 1550 nm wavelength, on \SVS crystals. In Figs.~\ref{1550}b,d we also include in dashed line the DC component of the reflectivity (in arbitrary units) measured simultaneously with the PKE, which reaffirms that the optical volume that we test indeed undergo the CDW transition.

\begin{figure}[h]
	\includegraphics[width=\columnwidth]{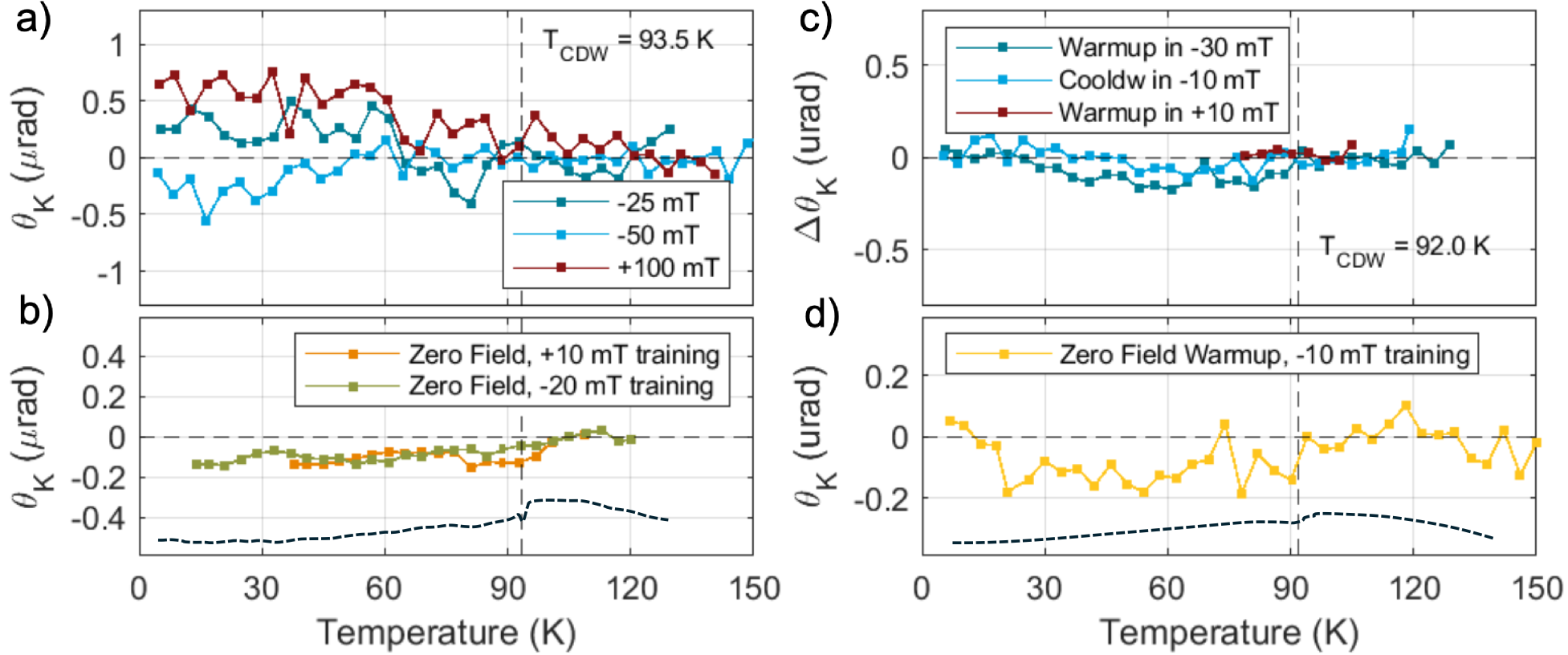}
	\caption{Polar Kerr Effect measurements at 830 nm wavelength: a,b)  on \CVS, and c,d) on \SVS. Similar to Fig.~\ref{1550}, measurements in magnetic field (panels (a and (c)) track the transition with signal of order of the respective Pauli susceptibility \cite{Saykin2023}. Zero-field warmup measurements after training the sample in a field while cooling it through the CDW transition (panels (a and (c)) Show no evidence for a spontaneous Kerr effect (see text). Dashed lines in panels (b) and (d) track the DC component of the reflectivity (in arbitrary units) measured simultaneously with the PKE, which reaffirms that the optical volume that we test indeed undergo the CDW transition.}
	\label{830}
\end{figure}

Following measurements using linearly polarized light at 800 nm \cite{Wu2021,Xu2022}, which reported a large Kerr response, up to 50 $\mu$rad, we used a 830 nm Sagnac system to search for such signals. Figs.~\ref{830}a,b show in-field and zero field data on \CVS while Figs.~\ref{830}c,d show data on \SVS crystals. While exhibiting somewhat larger error bars than when using the 1550 nm ZALSI, also here we see no evidence for spontaneous PKE for either kagom\'e  system. 

Comparing the 830 nm  to the 1550 nm results on \CVS, it is evident that the PKE response to a magnetic field at 1550 nm is several times larger than in 1550 nm, which is commensurate with optical measurements that identify a CDW-originated Lorentz peak at $\sim 0.7$ eV \cite{Uykur2021,Zhou2021}, with much breadth to include our 1550 nm measurements, but fades to no effect at $\sim 800$ nm. Signals at both, 1550 nm and 830 nm are further smaller for \SVS, again commensurate with optical studies on this material \cite{Hu2023}.

From the nature of the ZALSI as an ``absolute'' test for a finite spontaneous PKE (that is, no subtraction of signals and/or external field modulations are needed), we conclude that there is no spontaneous PKE in either \CVS or \SVS in the vicinity of either 1550 nm or 800 nm wavelength to within the instrument's resolution of $\sim 50$ nanorad, in contrast to measurements using analyses of reflected linearly polarized light \cite{Wu2021,Xu2022}. Our results also respond to a recent study of possible Kerr effect in \CVS \cite{Liu2025a} suggesting to explain the discrepancies between measurements at the two wavelengths. Here we further note that in general if a particular resonance at a frequency $\omega_r$ could be a ``best bet'' to observe the effect, then at the minimum we should be able to observe the tails of that resonance with a typical $(\omega_r/\omega_0)^2$ resolution. For example, a $\sim 1$ millirad signal due to a resonance at $\hbar\omega_r=1$ eV will be visible at $\hbar\omega_0=0.8$ eV (1550 nm) at a level of $\sim0.6$ millirad, or at $\sim0.4$ millirad at $\hbar\omega_0=1.5$ eV (830 nm), both much larger than our instrument's resolution.

We turn now to the study of Kerr effect under strain in \CVS. This material in particular was suggested to be sensitive to strain \cite{Guo2022,Guo2025}, which in turn require caution when samples are mounted for various measurements. Moreover, STM studies on the sister compound \RVS  \cite{Xing2024} showed evidence that the CDW relative intensity can be manipulated by inducing strain through  laser illumination with linearly polarized light along reciprocal lattice directions (which in real space will correspond to $b$ and $\bar{b}$). They further demonstrated that the same control can be achieved with a magnetic field applied in the $z$-direction, similar to previous reports on \CVS \cite{Jiang2021}. A simplified version of the analysis in \cite{Xing2024}, which displays coupling to both, strain and magnetic field is expected to be controlled by a two-components order parameter $\psi$, adding the following terms to free energy of the CDW system:
\begin{equation}
\Delta\mathcal{F}=-\textbf{m}\cdot\textbf{H}+\psi_1(\varepsilon_{xx}-\varepsilon_{yy})H_z+\psi_2\varepsilon_{xy}H_z
\end{equation}
where $\textbf{m}$ is the uniform magnetization if exists and the magnetic field is applied in the $z$-direction. With the directions $b$ and $\bar{b}$ equivalent, the second strain-field coupling term will induce a finite $\psi_2$ if we align the crystal along the $a$-direction. This in turn will result in a finite Kerr effect. The authors further suggested that residual strain in the $a$-direction induced in the crystals during experiments could result in a finite Kerr effect . While this is a plausible suggestion to explain small signals, it is an unlikely explanation for Kerr signals of order $\sim 50~\mu$rad (or larger) observed in experiments \cite{Wu2021,Xu2022}. Thus, expecting small signals, we searched for strain-induced Kerr signal using the Sagnac interferometer system.

\begin{figure}
    \centering
    \includegraphics[width=\linewidth]{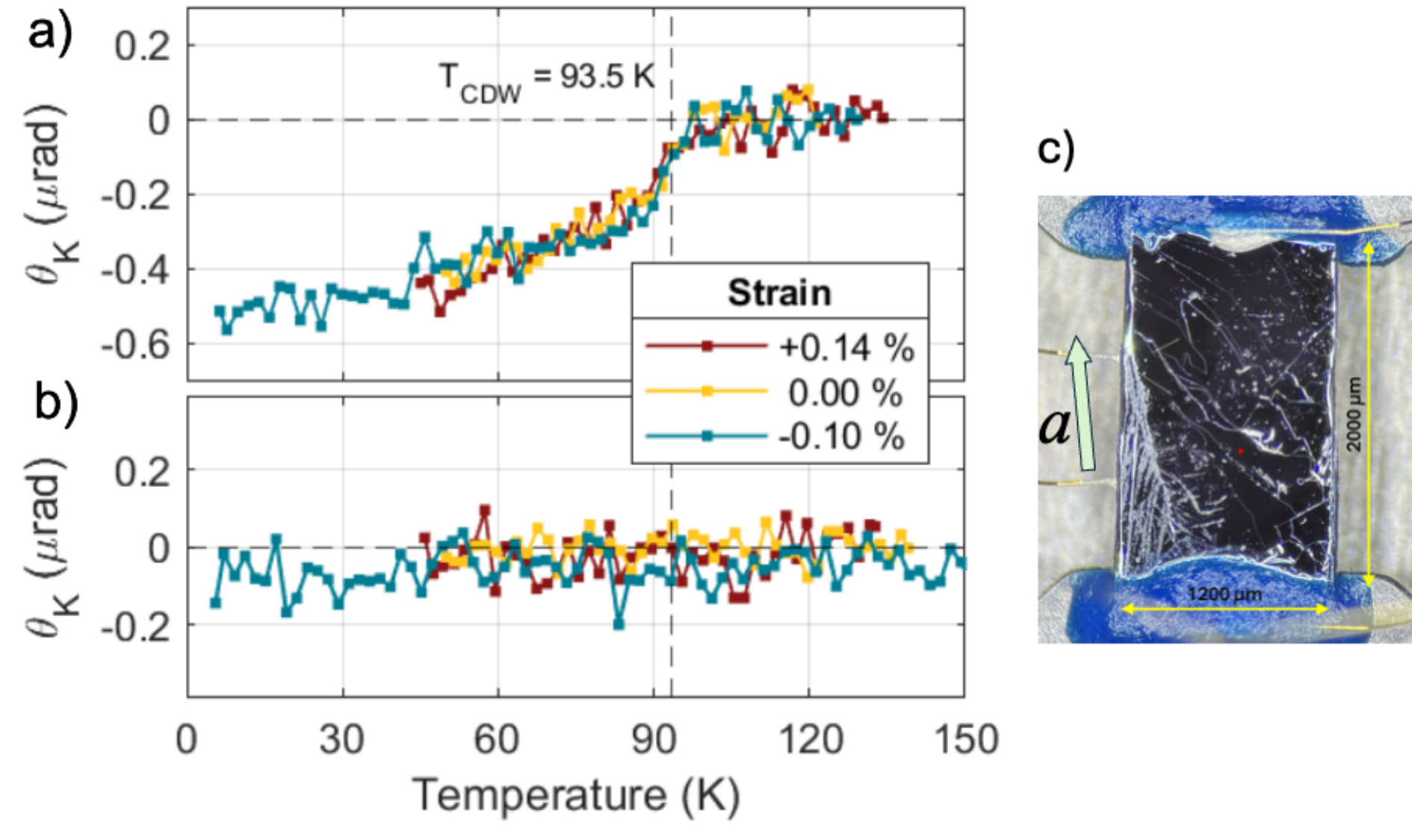}
    \caption{Kerr signal in strained \CVS at 1550 nm. a) Kerr effect measured in an applied field of 50 mT, yielding similar Kerr signal to previous in-field measurements. b) Zero-field warmup measurements after training in a field of 50 mT showing no discernible strain-induced signal through the CDW transition.   c) \CVS crystal cut and mounted onto the strain cell. Crystal dimensions 2mm$\times$1.2mm. Arrow marks the $a$-direction of the crystal in the cell (see Fig.~\ref{structure}.)}
    \label{CVS-1550-strain}
\end{figure}

Using the aforementioned ``Razorbill'' strain cell, we were able to apply uniaxial stress to the sample and cool it down through the CDW transition with tensile and compressive strains applied. While in practice the applied strain is not purely uniaxial, we could estimate the perpendicular contribution from the Poisson coefficients.  Either way we expect to see a finite signal which may change sign upon changing the strain direction. Figure~\ref{CVS-1550-strain} show Kerr effect results with either tensile or compressive strains applied to the $a$-direction (see Fig.~\ref{structure}.)
Analyzing the under-strain Kerr data we conclude that no discernible signal appears as a result of an applied uniaxial strain, either with an applied magnetic field or at zero field.

\section{Conclusions}
We report high resolution polar Kerr effect measurements on \CVS and \SVS single crystals in search for signatures of spontaneous polar Kerr effect (PKE) below the charge order transitions of these materials.  Utilizing two zero-area loop Sagnac interferometers operating at 1550 nm and 830 nm wavelengths, we studied the temperature dependence of possible PKE after training with magnetic field. While a finite field Kerr measurement yielded optical rotation expected from the Pauli susceptibility of the itinerant carriers, no signal was detected at zero-field  to within the noise floor limit of the apparatus at 30 nanoradians. Simultaneous  coherent reflection measurements  confirm the sharpness of the charge order transition in the same optical volume as the Kerr measurements.  Application of strain to reveal a hidden flux-ordered magnetic state did not result in a finite Kerr effect. We further note that our measurements were typically taken down to $\sim 10$ K, where no discernible signal could be detected at intermediate temperatures where we also observed anomalies in resistivity (See Supplementary Material \cite{supplement}.)

\section{Acknowledgements}
Work at Stanford University was supported by the U.S. Department of Energy, Office of Science, Basic Energy Sciences, Division of Materials Sciences and Engineering, under Contract DE-AC02-76SF00515.

\bibliography{kerr}

\begin{thebibliography}{52}%
\makeatletter
\providecommand \@ifxundefined [1]{%
 \@ifx{#1\undefined}
}%
\providecommand \@ifnum [1]{%
 \ifnum #1\expandafter \@firstoftwo
 \else \expandafter \@secondoftwo
 \fi
}%
\providecommand \@ifx [1]{%
 \ifx #1\expandafter \@firstoftwo
 \else \expandafter \@secondoftwo
 \fi
}%
\providecommand \natexlab [1]{#1}%
\providecommand \enquote  [1]{``#1''}%
\providecommand \bibnamefont  [1]{#1}%
\providecommand \bibfnamefont [1]{#1}%
\providecommand \citenamefont [1]{#1}%
\providecommand \href@noop [0]{\@secondoftwo}%
\providecommand \href [0]{\begingroup \@sanitize@url \@href}%
\providecommand \@href[1]{\@@startlink{#1}\@@href}%
\providecommand \@@href[1]{\endgroup#1\@@endlink}%
\providecommand \@sanitize@url [0]{\catcode `\\12\catcode `\$12\catcode
  `\&12\catcode `\#12\catcode `\^12\catcode `\_12\catcode `\%12\relax}%
\providecommand \@@startlink[1]{}%
\providecommand \@@endlink[0]{}%
\providecommand \url  [0]{\begingroup\@sanitize@url \@url }%
\providecommand \@url [1]{\endgroup\@href {#1}{\urlprefix }}%
\providecommand \urlprefix  [0]{URL }%
\providecommand \Eprint [0]{\href }%
\providecommand \doibase [0]{https://doi.org/}%
\providecommand \selectlanguage [0]{\@gobble}%
\providecommand \bibinfo  [0]{\@secondoftwo}%
\providecommand \bibfield  [0]{\@secondoftwo}%
\providecommand \translation [1]{[#1]}%
\providecommand \BibitemOpen [0]{}%
\providecommand \bibitemStop [0]{}%
\providecommand \bibitemNoStop [0]{.\EOS\space}%
\providecommand \EOS [0]{\spacefactor3000\relax}%
\providecommand \BibitemShut  [1]{\csname bibitem#1\endcsname}%
\let\auto@bib@innerbib\@empty
\bibitem [{\citenamefont {Wilson}\ and\ \citenamefont
  {Ortiz}(2024)}]{Wilson2024}%
  \BibitemOpen
  \bibfield  {author} {\bibinfo {author} {\bibfnamefont {S.~D.}\ \bibnamefont
  {Wilson}}\ and\ \bibinfo {author} {\bibfnamefont {B.~R.}\ \bibnamefont
  {Ortiz}},\ }\bibfield  {title} {\bibinfo {title} {{AV$_3$Sb$_5$ kagome
  superconductors}},\ }\href {https://doi.org/10.1038/s41578-024-00677-y}
  {\bibfield  {journal} {\bibinfo  {journal} {Nature Reviews Materials}\
  }\textbf {\bibinfo {volume} {9}},\ \bibinfo {pages} {420} (\bibinfo {year}
  {2024})}\BibitemShut {NoStop}%
\bibitem [{\citenamefont {Wang}\ \emph {et~al.}(2025)\citenamefont {Wang},
  \citenamefont {Lei}, \citenamefont {Qi},\ and\ \citenamefont
  {Felser}}]{Wang2025}%
  \BibitemOpen
  \bibfield  {author} {\bibinfo {author} {\bibfnamefont {Q.}~\bibnamefont
  {Wang}}, \bibinfo {author} {\bibfnamefont {H.}~\bibnamefont {Lei}}, \bibinfo
  {author} {\bibfnamefont {Y.}~\bibnamefont {Qi}},\ and\ \bibinfo {author}
  {\bibfnamefont {C.}~\bibnamefont {Felser}},\ }\bibfield  {title} {\bibinfo
  {title} {Intriguing kagome topological materials},\ }\href
  {https://doi.org/10.1038/s41535-025-00790-3} {\bibfield  {journal} {\bibinfo
  {journal} {npj Quantum Materials}\ }\textbf {\bibinfo {volume} {10}},\
  \bibinfo {pages} {72} (\bibinfo {year} {2025})}\BibitemShut {NoStop}%
\bibitem [{\citenamefont {Li}\ \emph {et~al.}(2025)\citenamefont {Li},
  \citenamefont {Ma}, \citenamefont {Lou},\ and\ \citenamefont
  {Wang}}]{Li2025}%
  \BibitemOpen
  \bibfield  {author} {\bibinfo {author} {\bibfnamefont {M.}~\bibnamefont
  {Li}}, \bibinfo {author} {\bibfnamefont {H.}~\bibnamefont {Ma}}, \bibinfo
  {author} {\bibfnamefont {R.}~\bibnamefont {Lou}},\ and\ \bibinfo {author}
  {\bibfnamefont {S.}~\bibnamefont {Wang}},\ }\bibfield  {title} {\bibinfo
  {title} {Electronic band structures of topological kagome materials},\ }\href
  {https://doi.org/10.1088/1674-1056/ad925d} {\bibfield  {journal} {\bibinfo
  {journal} {Chinese Physics B}\ }\textbf {\bibinfo {volume} {34}},\ \bibinfo
  {pages} {017101} (\bibinfo {year} {2025})}\BibitemShut {NoStop}%
\bibitem [{\citenamefont {Ortiz}\ \emph {et~al.}(2020)\citenamefont {Ortiz},
  \citenamefont {Teicher}, \citenamefont {Hu}, \citenamefont {Zuo},
  \citenamefont {Sarte}, \citenamefont {Schueller}, \citenamefont {Abeykoon},
  \citenamefont {Krogstad}, \citenamefont {Rosenkranz}, \citenamefont {Osborn},
  \citenamefont {Seshadri}, \citenamefont {Balents}, \citenamefont {He},\ and\
  \citenamefont {Wilson}}]{Ortiz2020}%
  \BibitemOpen
  \bibfield  {author} {\bibinfo {author} {\bibfnamefont {B.~R.}\ \bibnamefont
  {Ortiz}}, \bibinfo {author} {\bibfnamefont {S.~M.~L.}\ \bibnamefont
  {Teicher}}, \bibinfo {author} {\bibfnamefont {Y.}~\bibnamefont {Hu}},
  \bibinfo {author} {\bibfnamefont {J.~L.}\ \bibnamefont {Zuo}}, \bibinfo
  {author} {\bibfnamefont {P.~M.}\ \bibnamefont {Sarte}}, \bibinfo {author}
  {\bibfnamefont {E.~C.}\ \bibnamefont {Schueller}}, \bibinfo {author}
  {\bibfnamefont {A.~M.~M.}\ \bibnamefont {Abeykoon}}, \bibinfo {author}
  {\bibfnamefont {M.~J.}\ \bibnamefont {Krogstad}}, \bibinfo {author}
  {\bibfnamefont {S.}~\bibnamefont {Rosenkranz}}, \bibinfo {author}
  {\bibfnamefont {R.}~\bibnamefont {Osborn}}, \bibinfo {author} {\bibfnamefont
  {R.}~\bibnamefont {Seshadri}}, \bibinfo {author} {\bibfnamefont
  {L.}~\bibnamefont {Balents}}, \bibinfo {author} {\bibfnamefont
  {J.}~\bibnamefont {He}},\ and\ \bibinfo {author} {\bibfnamefont {S.~D.}\
  \bibnamefont {Wilson}},\ }\bibfield  {title} {\bibinfo {title}
  {{CsV$_3$Sb$_5$}: A ${\mathbb{z}}_{2}$ topological kagom\'e metal with a
  superconducting ground state},\ }\href
  {https://doi.org/10.1103/PhysRevLett.125.247002} {\bibfield  {journal}
  {\bibinfo  {journal} {Phys. Rev. Lett.}\ }\textbf {\bibinfo {volume} {125}},\
  \bibinfo {pages} {247002} (\bibinfo {year} {2020})}\BibitemShut {NoStop}%
\bibitem [{\citenamefont {Arachchige}\ \emph {et~al.}(2022)\citenamefont
  {Arachchige}, \citenamefont {Meier}, \citenamefont {Marshall}, \citenamefont
  {Matsuoka}, \citenamefont {Xue}, \citenamefont {McGuire}, \citenamefont
  {Hermann}, \citenamefont {Cao},\ and\ \citenamefont
  {Mandrus}}]{Arachchige2022}%
  \BibitemOpen
  \bibfield  {author} {\bibinfo {author} {\bibfnamefont {H.~W.~S.}\
  \bibnamefont {Arachchige}}, \bibinfo {author} {\bibfnamefont {W.~R.}\
  \bibnamefont {Meier}}, \bibinfo {author} {\bibfnamefont {M.}~\bibnamefont
  {Marshall}}, \bibinfo {author} {\bibfnamefont {T.}~\bibnamefont {Matsuoka}},
  \bibinfo {author} {\bibfnamefont {R.}~\bibnamefont {Xue}}, \bibinfo {author}
  {\bibfnamefont {M.~A.}\ \bibnamefont {McGuire}}, \bibinfo {author}
  {\bibfnamefont {R.~P.}\ \bibnamefont {Hermann}}, \bibinfo {author}
  {\bibfnamefont {H.}~\bibnamefont {Cao}},\ and\ \bibinfo {author}
  {\bibfnamefont {D.}~\bibnamefont {Mandrus}},\ }\bibfield  {title} {\bibinfo
  {title} {Charge density wave in kagome lattice intermetallic
  {ScV$_6$Sn$_6$}},\ }\href@noop {} {\bibfield  {journal} {\bibinfo  {journal}
  {Physical Review Letters}\ }\textbf {\bibinfo {volume} {129}},\ \bibinfo
  {pages} {216402} (\bibinfo {year} {2022})}\BibitemShut {NoStop}%
\bibitem [{\citenamefont {Feng}\ \emph {et~al.}(2021)\citenamefont {Feng},
  \citenamefont {Jiang}, \citenamefont {Wang},\ and\ \citenamefont
  {Hu}}]{Feng2021}%
  \BibitemOpen
  \bibfield  {author} {\bibinfo {author} {\bibfnamefont {X.}~\bibnamefont
  {Feng}}, \bibinfo {author} {\bibfnamefont {K.}~\bibnamefont {Jiang}},
  \bibinfo {author} {\bibfnamefont {Z.}~\bibnamefont {Wang}},\ and\ \bibinfo
  {author} {\bibfnamefont {J.}~\bibnamefont {Hu}},\ }\bibfield  {title}
  {\bibinfo {title} {Chiral flux phase in the kagom\'e superconductor
  {AV$_3$Sb$_5$}},\ }\href
  {https://doi.org/https://doi.org/10.1016/j.scib.2021.04.043} {\bibfield
  {journal} {\bibinfo  {journal} {Science Bulletin}\ }\textbf {\bibinfo
  {volume} {66}},\ \bibinfo {pages} {1384} (\bibinfo {year}
  {2021})}\BibitemShut {NoStop}%
\bibitem [{\citenamefont {Denner}\ \emph {et~al.}(2021)\citenamefont {Denner},
  \citenamefont {Thomale},\ and\ \citenamefont {Neupert}}]{Denner2021}%
  \BibitemOpen
  \bibfield  {author} {\bibinfo {author} {\bibfnamefont {M.~M.}\ \bibnamefont
  {Denner}}, \bibinfo {author} {\bibfnamefont {R.}~\bibnamefont {Thomale}},\
  and\ \bibinfo {author} {\bibfnamefont {T.}~\bibnamefont {Neupert}},\
  }\bibfield  {title} {\bibinfo {title} {Analysis of charge order in the kagome
  metal {$A{\mathrm{V}}_{3}{\mathrm{Sb}}_{5}$
  ($A=\mathrm{K},\mathrm{Rb},\mathrm{Cs}$)}},\ }\href
  {https://doi.org/10.1103/PhysRevLett.127.217601} {\bibfield  {journal}
  {\bibinfo  {journal} {Phys. Rev. Lett.}\ }\textbf {\bibinfo {volume} {127}},\
  \bibinfo {pages} {217601} (\bibinfo {year} {2021})}\BibitemShut {NoStop}%
\bibitem [{\citenamefont {Li}\ \emph {et~al.}(2024)\citenamefont {Li},
  \citenamefont {Kim},\ and\ \citenamefont {Kee}}]{Li2024}%
  \BibitemOpen
  \bibfield  {author} {\bibinfo {author} {\bibfnamefont {H.}~\bibnamefont
  {Li}}, \bibinfo {author} {\bibfnamefont {Y.~B.}\ \bibnamefont {Kim}},\ and\
  \bibinfo {author} {\bibfnamefont {H.-Y.}\ \bibnamefont {Kee}},\ }\bibfield
  {title} {\bibinfo {title} {Intertwined van hove singularities as a mechanism
  for loop current order in kagome metals},\ }\href
  {https://doi.org/10.1103/PhysRevLett.132.146501} {\bibfield  {journal}
  {\bibinfo  {journal} {Phys. Rev. Lett.}\ }\textbf {\bibinfo {volume} {132}},\
  \bibinfo {pages} {146501} (\bibinfo {year} {2024})}\BibitemShut {NoStop}%
\bibitem [{\citenamefont {Guo}\ \emph {et~al.}(2022)\citenamefont {Guo},
  \citenamefont {Putzke}, \citenamefont {Konyzheva}, \citenamefont {Huang},
  \citenamefont {Gutierrez-Amigo}, \citenamefont {Errea}, \citenamefont {Chen},
  \citenamefont {Vergniory}, \citenamefont {Felser}, \citenamefont {Fischer},
  \citenamefont {Neupert},\ and\ \citenamefont {Moll}}]{Guo2022}%
  \BibitemOpen
  \bibfield  {author} {\bibinfo {author} {\bibfnamefont {C.}~\bibnamefont
  {Guo}}, \bibinfo {author} {\bibfnamefont {C.}~\bibnamefont {Putzke}},
  \bibinfo {author} {\bibfnamefont {S.}~\bibnamefont {Konyzheva}}, \bibinfo
  {author} {\bibfnamefont {X.}~\bibnamefont {Huang}}, \bibinfo {author}
  {\bibfnamefont {M.}~\bibnamefont {Gutierrez-Amigo}}, \bibinfo {author}
  {\bibfnamefont {I.}~\bibnamefont {Errea}}, \bibinfo {author} {\bibfnamefont
  {D.}~\bibnamefont {Chen}}, \bibinfo {author} {\bibfnamefont {M.~G.}\
  \bibnamefont {Vergniory}}, \bibinfo {author} {\bibfnamefont {C.}~\bibnamefont
  {Felser}}, \bibinfo {author} {\bibfnamefont {M.~H.}\ \bibnamefont {Fischer}},
  \bibinfo {author} {\bibfnamefont {T.}~\bibnamefont {Neupert}},\ and\ \bibinfo
  {author} {\bibfnamefont {P.~J.~W.}\ \bibnamefont {Moll}},\ }\bibfield
  {title} {\bibinfo {title} {Switchable chiral transport in charge-ordered
  kagome metal {CsV$_3$Sb$_5$}},\ }\href
  {https://doi.org/10.1038/s41586-022-05127-9} {\bibfield  {journal} {\bibinfo
  {journal} {Nature}\ }\textbf {\bibinfo {volume} {611}},\ \bibinfo {pages}
  {461} (\bibinfo {year} {2022})}\BibitemShut {NoStop}%
\bibitem [{\citenamefont {{C. Guo, K. Wang, L. Zhang, C. Putzke, D. Chen, M. R.
  van Delft, S. Wiedmann, F. F. Balakirev, R. D. McDonald, M. Gutierrez-Amigo,
  M. Alkorta, I. Errea, M. G. Vergniory, T. Oka, R. Moessner, M. H. Fischer, T.
  Neupert, C. Felser, P. J.W. Moll}}(2025)}]{Guo2025}%
  \BibitemOpen
  \bibfield  {author} {\bibinfo {author} {\bibnamefont {{C. Guo, K. Wang, L.
  Zhang, C. Putzke, D. Chen, M. R. van Delft, S. Wiedmann, F. F. Balakirev, R.
  D. McDonald, M. Gutierrez-Amigo, M. Alkorta, I. Errea, M. G. Vergniory, T.
  Oka, R. Moessner, M. H. Fischer, T. Neupert, C. Felser, P. J.W. Moll}}},\
  }\href@noop {} {\bibinfo {title} {Long-range electron coherence in kagome
  metals}} (\bibinfo {year} {2025}),\ \Eprint
  {https://arxiv.org/abs/2504.13564} {arXiv:2504.13564 [cond-mat.str-el]}
  \BibitemShut {NoStop}%
\bibitem [{\citenamefont {Alkorta}\ \emph {et~al.}(2025)\citenamefont
  {Alkorta}, \citenamefont {Gutierrez-Amigo}, \citenamefont {Guo},
  \citenamefont {Moll}, \citenamefont {Vergniory},\ and\ \citenamefont
  {Errea.}}]{Alkorta2025}%
  \BibitemOpen
  \bibfield  {author} {\bibinfo {author} {\bibfnamefont {M.}~\bibnamefont
  {Alkorta}}, \bibinfo {author} {\bibfnamefont {M.}~\bibnamefont
  {Gutierrez-Amigo}}, \bibinfo {author} {\bibfnamefont {C.}~\bibnamefont
  {Guo}}, \bibinfo {author} {\bibfnamefont {P.~J.~W.}\ \bibnamefont {Moll}},
  \bibinfo {author} {\bibfnamefont {M.~G.}\ \bibnamefont {Vergniory}},\ and\
  \bibinfo {author} {\bibfnamefont {I.}~\bibnamefont {Errea.}},\ }\href@noop {}
  {\bibinfo {title} {Symmetry-broken charge-ordered ground state in
  {CsV$_3$Sb$_5$} kagome metal}} (\bibinfo {year} {2025}),\ \Eprint
  {https://arxiv.org/abs/2505.19686v2} {arXiv:2505.19686v2 [cond-mat.mtrl-sci]}
  \BibitemShut {NoStop}%
\bibitem [{\citenamefont {Wang}\ \emph {et~al.}(2021)\citenamefont {Wang},
  \citenamefont {Jiang}, \citenamefont {Yin}, \citenamefont {Li}, \citenamefont
  {Wang}, \citenamefont {Huang}, \citenamefont {Shao}, \citenamefont {Liu},
  \citenamefont {Zhu}, \citenamefont {Shumiya}, \citenamefont {Hossain},
  \citenamefont {Liu}, \citenamefont {Shi}, \citenamefont {Duan}, \citenamefont
  {Li}, \citenamefont {Chang}, \citenamefont {Dai}, \citenamefont {Ye},
  \citenamefont {Xu}, \citenamefont {Wang}, \citenamefont {Zheng},
  \citenamefont {Jia}, \citenamefont {Hasan},\ and\ \citenamefont
  {Yao}}]{Wang2021}%
  \BibitemOpen
  \bibfield  {author} {\bibinfo {author} {\bibfnamefont {Z.}~\bibnamefont
  {Wang}}, \bibinfo {author} {\bibfnamefont {Y.-X.}\ \bibnamefont {Jiang}},
  \bibinfo {author} {\bibfnamefont {J.-X.}\ \bibnamefont {Yin}}, \bibinfo
  {author} {\bibfnamefont {Y.}~\bibnamefont {Li}}, \bibinfo {author}
  {\bibfnamefont {G.-Y.}\ \bibnamefont {Wang}}, \bibinfo {author}
  {\bibfnamefont {H.-L.}\ \bibnamefont {Huang}}, \bibinfo {author}
  {\bibfnamefont {S.}~\bibnamefont {Shao}}, \bibinfo {author} {\bibfnamefont
  {J.}~\bibnamefont {Liu}}, \bibinfo {author} {\bibfnamefont {P.}~\bibnamefont
  {Zhu}}, \bibinfo {author} {\bibfnamefont {N.}~\bibnamefont {Shumiya}},
  \bibinfo {author} {\bibfnamefont {M.~S.}\ \bibnamefont {Hossain}}, \bibinfo
  {author} {\bibfnamefont {H.}~\bibnamefont {Liu}}, \bibinfo {author}
  {\bibfnamefont {Y.}~\bibnamefont {Shi}}, \bibinfo {author} {\bibfnamefont
  {J.}~\bibnamefont {Duan}}, \bibinfo {author} {\bibfnamefont {X.}~\bibnamefont
  {Li}}, \bibinfo {author} {\bibfnamefont {G.}~\bibnamefont {Chang}}, \bibinfo
  {author} {\bibfnamefont {P.}~\bibnamefont {Dai}}, \bibinfo {author}
  {\bibfnamefont {Z.}~\bibnamefont {Ye}}, \bibinfo {author} {\bibfnamefont
  {G.}~\bibnamefont {Xu}}, \bibinfo {author} {\bibfnamefont {Y.}~\bibnamefont
  {Wang}}, \bibinfo {author} {\bibfnamefont {H.}~\bibnamefont {Zheng}},
  \bibinfo {author} {\bibfnamefont {J.}~\bibnamefont {Jia}}, \bibinfo {author}
  {\bibfnamefont {M.~Z.}\ \bibnamefont {Hasan}},\ and\ \bibinfo {author}
  {\bibfnamefont {Y.}~\bibnamefont {Yao}},\ }\bibfield  {title} {\bibinfo
  {title} {Electronic nature of chiral charge order in the kagom\'e
  superconductor {CsV$_3$Sb$_5$}},\ }\href
  {https://doi.org/10.1103/PhysRevB.104.075148} {\bibfield  {journal} {\bibinfo
   {journal} {Phys. Rev. B}\ }\textbf {\bibinfo {volume} {104}},\ \bibinfo
  {pages} {075148} (\bibinfo {year} {2021})}\BibitemShut {NoStop}%
\bibitem [{\citenamefont {Jiang}\ \emph {et~al.}(2021)\citenamefont {Jiang},
  \citenamefont {Yin}, \citenamefont {Denner}, \citenamefont {Shumiya},
  \citenamefont {Ortiz}, \citenamefont {Xu}, \citenamefont {Guguchia},
  \citenamefont {He}, \citenamefont {Hossain}, \citenamefont {Liu},
  \citenamefont {Ruff}, \citenamefont {Kautzsch}, \citenamefont {Zhang},
  \citenamefont {Chang}, \citenamefont {Belopolski}, \citenamefont {Zhang},
  \citenamefont {Cochran}, \citenamefont {Multer}, \citenamefont {Litskevich},
  \citenamefont {Cheng}, \citenamefont {Yang}, \citenamefont {Wang},
  \citenamefont {Thomale}, \citenamefont {Neupert}, \citenamefont {Wilson},\
  and\ \citenamefont {Hasan}}]{Jiang2021}%
  \BibitemOpen
  \bibfield  {author} {\bibinfo {author} {\bibfnamefont {Y.-X.}\ \bibnamefont
  {Jiang}}, \bibinfo {author} {\bibfnamefont {J.-X.}\ \bibnamefont {Yin}},
  \bibinfo {author} {\bibfnamefont {M.~M.}\ \bibnamefont {Denner}}, \bibinfo
  {author} {\bibfnamefont {N.}~\bibnamefont {Shumiya}}, \bibinfo {author}
  {\bibfnamefont {B.~R.}\ \bibnamefont {Ortiz}}, \bibinfo {author}
  {\bibfnamefont {G.}~\bibnamefont {Xu}}, \bibinfo {author} {\bibfnamefont
  {Z.}~\bibnamefont {Guguchia}}, \bibinfo {author} {\bibfnamefont
  {J.}~\bibnamefont {He}}, \bibinfo {author} {\bibfnamefont {M.~S.}\
  \bibnamefont {Hossain}}, \bibinfo {author} {\bibfnamefont {X.}~\bibnamefont
  {Liu}}, \bibinfo {author} {\bibfnamefont {J.}~\bibnamefont {Ruff}}, \bibinfo
  {author} {\bibfnamefont {L.}~\bibnamefont {Kautzsch}}, \bibinfo {author}
  {\bibfnamefont {S.~S.}\ \bibnamefont {Zhang}}, \bibinfo {author}
  {\bibfnamefont {G.}~\bibnamefont {Chang}}, \bibinfo {author} {\bibfnamefont
  {I.}~\bibnamefont {Belopolski}}, \bibinfo {author} {\bibfnamefont
  {Q.}~\bibnamefont {Zhang}}, \bibinfo {author} {\bibfnamefont {T.~A.}\
  \bibnamefont {Cochran}}, \bibinfo {author} {\bibfnamefont {D.}~\bibnamefont
  {Multer}}, \bibinfo {author} {\bibfnamefont {M.}~\bibnamefont {Litskevich}},
  \bibinfo {author} {\bibfnamefont {Z.-J.}\ \bibnamefont {Cheng}}, \bibinfo
  {author} {\bibfnamefont {X.~P.}\ \bibnamefont {Yang}}, \bibinfo {author}
  {\bibfnamefont {Z.}~\bibnamefont {Wang}}, \bibinfo {author} {\bibfnamefont
  {R.}~\bibnamefont {Thomale}}, \bibinfo {author} {\bibfnamefont
  {T.}~\bibnamefont {Neupert}}, \bibinfo {author} {\bibfnamefont {S.~D.}\
  \bibnamefont {Wilson}},\ and\ \bibinfo {author} {\bibfnamefont {M.~Z.}\
  \bibnamefont {Hasan}},\ }\bibfield  {title} {\bibinfo {title} {Unconventional
  chiral charge order in kagom\'e superconductor {KV$_3$Sb$_5$}},\ }\href
  {https://doi.org/10.1038/s41563-021-01034-y} {\bibfield  {journal} {\bibinfo
  {journal} {Nature Materials}\ }\textbf {\bibinfo {volume} {20}},\ \bibinfo
  {pages} {1353} (\bibinfo {year} {2021})}\BibitemShut {NoStop}%
\bibitem [{\citenamefont {Shumiya}\ \emph {et~al.}(2021)\citenamefont
  {Shumiya}, \citenamefont {Hossain}, \citenamefont {Yin}, \citenamefont
  {Jiang}, \citenamefont {Ortiz}, \citenamefont {Liu}, \citenamefont {Shi},
  \citenamefont {Yin}, \citenamefont {Lei}, \citenamefont {Zhang},
  \citenamefont {Chang}, \citenamefont {Zhang}, \citenamefont {Cochran},
  \citenamefont {Multer}, \citenamefont {Litskevich}, \citenamefont {Cheng},
  \citenamefont {Yang}, \citenamefont {Guguchia}, \citenamefont {Wilson},\ and\
  \citenamefont {Hasan}}]{Shumiya2021}%
  \BibitemOpen
  \bibfield  {author} {\bibinfo {author} {\bibfnamefont {N.}~\bibnamefont
  {Shumiya}}, \bibinfo {author} {\bibfnamefont {M.~S.}\ \bibnamefont
  {Hossain}}, \bibinfo {author} {\bibfnamefont {J.-X.}\ \bibnamefont {Yin}},
  \bibinfo {author} {\bibfnamefont {Y.-X.}\ \bibnamefont {Jiang}}, \bibinfo
  {author} {\bibfnamefont {B.~R.}\ \bibnamefont {Ortiz}}, \bibinfo {author}
  {\bibfnamefont {H.}~\bibnamefont {Liu}}, \bibinfo {author} {\bibfnamefont
  {Y.}~\bibnamefont {Shi}}, \bibinfo {author} {\bibfnamefont {Q.}~\bibnamefont
  {Yin}}, \bibinfo {author} {\bibfnamefont {H.}~\bibnamefont {Lei}}, \bibinfo
  {author} {\bibfnamefont {S.~S.}\ \bibnamefont {Zhang}}, \bibinfo {author}
  {\bibfnamefont {G.}~\bibnamefont {Chang}}, \bibinfo {author} {\bibfnamefont
  {Q.}~\bibnamefont {Zhang}}, \bibinfo {author} {\bibfnamefont {T.~A.}\
  \bibnamefont {Cochran}}, \bibinfo {author} {\bibfnamefont {D.}~\bibnamefont
  {Multer}}, \bibinfo {author} {\bibfnamefont {M.}~\bibnamefont {Litskevich}},
  \bibinfo {author} {\bibfnamefont {Z.-J.}\ \bibnamefont {Cheng}}, \bibinfo
  {author} {\bibfnamefont {X.~P.}\ \bibnamefont {Yang}}, \bibinfo {author}
  {\bibfnamefont {Z.}~\bibnamefont {Guguchia}}, \bibinfo {author}
  {\bibfnamefont {S.~D.}\ \bibnamefont {Wilson}},\ and\ \bibinfo {author}
  {\bibfnamefont {M.~Z.}\ \bibnamefont {Hasan}},\ }\bibfield  {title} {\bibinfo
  {title} {Intrinsic nature of chiral charge order in the kagom\'e
  superconductor {RbV$_3$Sb$_5$}},\ }\href
  {https://doi.org/10.1103/PhysRevB.104.035131} {\bibfield  {journal} {\bibinfo
   {journal} {Phys. Rev. B}\ }\textbf {\bibinfo {volume} {104}},\ \bibinfo
  {pages} {035131} (\bibinfo {year} {2021})}\BibitemShut {NoStop}%
\bibitem [{\citenamefont {Li}\ \emph {et~al.}(2022{\natexlab{a}})\citenamefont
  {Li}, \citenamefont {Wan}, \citenamefont {Li}, \citenamefont {Li},
  \citenamefont {Gu}, \citenamefont {Yang}, \citenamefont {Li}, \citenamefont
  {Wang}, \citenamefont {Yao},\ and\ \citenamefont {Wen}}]{Li2022CVS}%
  \BibitemOpen
  \bibfield  {author} {\bibinfo {author} {\bibfnamefont {H.}~\bibnamefont
  {Li}}, \bibinfo {author} {\bibfnamefont {S.}~\bibnamefont {Wan}}, \bibinfo
  {author} {\bibfnamefont {H.}~\bibnamefont {Li}}, \bibinfo {author}
  {\bibfnamefont {Q.}~\bibnamefont {Li}}, \bibinfo {author} {\bibfnamefont
  {Q.}~\bibnamefont {Gu}}, \bibinfo {author} {\bibfnamefont {H.}~\bibnamefont
  {Yang}}, \bibinfo {author} {\bibfnamefont {Y.}~\bibnamefont {Li}}, \bibinfo
  {author} {\bibfnamefont {Z.}~\bibnamefont {Wang}}, \bibinfo {author}
  {\bibfnamefont {Y.}~\bibnamefont {Yao}},\ and\ \bibinfo {author}
  {\bibfnamefont {H.-H.}\ \bibnamefont {Wen}},\ }\bibfield  {title} {\bibinfo
  {title} {No observation of chiral flux current in the topological kagome
  metal {${\mathrm{CsV}}_{3}{\mathrm{Sb}}_{5}$}},\ }\href
  {https://doi.org/10.1103/PhysRevB.105.045102} {\bibfield  {journal} {\bibinfo
   {journal} {Phys. Rev. B}\ }\textbf {\bibinfo {volume} {105}},\ \bibinfo
  {pages} {045102} (\bibinfo {year} {2022}{\natexlab{a}})}\BibitemShut
  {NoStop}%
\bibitem [{\citenamefont {Li}\ \emph {et~al.}(2022{\natexlab{b}})\citenamefont
  {Li}, \citenamefont {Zhao}, \citenamefont {Ortiz}, \citenamefont {Park},
  \citenamefont {Ye}, \citenamefont {Balents}, \citenamefont {Wang},
  \citenamefont {Wilson},\ and\ \citenamefont {Zeljkovic}}]{Li2022KVS}%
  \BibitemOpen
  \bibfield  {author} {\bibinfo {author} {\bibfnamefont {H.}~\bibnamefont
  {Li}}, \bibinfo {author} {\bibfnamefont {H.}~\bibnamefont {Zhao}}, \bibinfo
  {author} {\bibfnamefont {B.~R.}\ \bibnamefont {Ortiz}}, \bibinfo {author}
  {\bibfnamefont {T.}~\bibnamefont {Park}}, \bibinfo {author} {\bibfnamefont
  {M.}~\bibnamefont {Ye}}, \bibinfo {author} {\bibfnamefont {L.}~\bibnamefont
  {Balents}}, \bibinfo {author} {\bibfnamefont {Z.}~\bibnamefont {Wang}},
  \bibinfo {author} {\bibfnamefont {S.~D.}\ \bibnamefont {Wilson}},\ and\
  \bibinfo {author} {\bibfnamefont {I.}~\bibnamefont {Zeljkovic}},\ }\bibfield
  {title} {\bibinfo {title} {Rotation symmetry breaking in the normal state of
  a kagome superconductor {KV$_3$Sb$_5$}},\ }\href
  {https://doi.org/10.1038/s41567-021-01479-7} {\bibfield  {journal} {\bibinfo
  {journal} {Nature Physics}\ }\textbf {\bibinfo {volume} {18}},\ \bibinfo
  {pages} {265} (\bibinfo {year} {2022}{\natexlab{b}})}\BibitemShut {NoStop}%
\bibitem [{\citenamefont {Xing}\ \emph {et~al.}(2024)\citenamefont {Xing},
  \citenamefont {Bae}, \citenamefont {Ritz}, \citenamefont {Yang},
  \citenamefont {Birol}, \citenamefont {Capa~Salinas}, \citenamefont {Ortiz},
  \citenamefont {Wilson}, \citenamefont {Wang}, \citenamefont {Fernandes},\
  and\ \citenamefont {Madhavan}}]{Xing2024}%
  \BibitemOpen
  \bibfield  {author} {\bibinfo {author} {\bibfnamefont {Y.}~\bibnamefont
  {Xing}}, \bibinfo {author} {\bibfnamefont {S.}~\bibnamefont {Bae}}, \bibinfo
  {author} {\bibfnamefont {E.}~\bibnamefont {Ritz}}, \bibinfo {author}
  {\bibfnamefont {F.}~\bibnamefont {Yang}}, \bibinfo {author} {\bibfnamefont
  {T.}~\bibnamefont {Birol}}, \bibinfo {author} {\bibfnamefont {A.~N.}\
  \bibnamefont {Capa~Salinas}}, \bibinfo {author} {\bibfnamefont {B.~R.}\
  \bibnamefont {Ortiz}}, \bibinfo {author} {\bibfnamefont {S.~D.}\ \bibnamefont
  {Wilson}}, \bibinfo {author} {\bibfnamefont {Z.}~\bibnamefont {Wang}},
  \bibinfo {author} {\bibfnamefont {R.~M.}\ \bibnamefont {Fernandes}},\ and\
  \bibinfo {author} {\bibfnamefont {V.}~\bibnamefont {Madhavan}},\ }\bibfield
  {title} {\bibinfo {title} {Optical manipulation of the charge-density-wave
  state in {RbV$_3$Sb$_5$}},\ }\href
  {https://doi.org/10.1038/s41586-024-07519-5} {\bibfield  {journal} {\bibinfo
  {journal} {Nature}\ }\textbf {\bibinfo {volume} {631}},\ \bibinfo {pages}
  {60} (\bibinfo {year} {2024})}\BibitemShut {NoStop}%
\bibitem [{\citenamefont {Khasanov}\ \emph {et~al.}(2022)\citenamefont
  {Khasanov}, \citenamefont {Das}, \citenamefont {Gupta}, \citenamefont
  {Mielke}, \citenamefont {Elender}, \citenamefont {Yin}, \citenamefont {Tu},
  \citenamefont {Gong}, \citenamefont {Lei}, \citenamefont {Ritz},
  \citenamefont {Fernandes}, \citenamefont {Birol}, \citenamefont {Guguchia},\
  and\ \citenamefont {Luetkens}}]{Khasanov2022}%
  \BibitemOpen
  \bibfield  {author} {\bibinfo {author} {\bibfnamefont {R.}~\bibnamefont
  {Khasanov}}, \bibinfo {author} {\bibfnamefont {D.}~\bibnamefont {Das}},
  \bibinfo {author} {\bibfnamefont {R.}~\bibnamefont {Gupta}}, \bibinfo
  {author} {\bibfnamefont {C.}~\bibnamefont {Mielke}}, \bibinfo {author}
  {\bibfnamefont {M.}~\bibnamefont {Elender}}, \bibinfo {author} {\bibfnamefont
  {Q.}~\bibnamefont {Yin}}, \bibinfo {author} {\bibfnamefont {Z.}~\bibnamefont
  {Tu}}, \bibinfo {author} {\bibfnamefont {C.}~\bibnamefont {Gong}}, \bibinfo
  {author} {\bibfnamefont {H.}~\bibnamefont {Lei}}, \bibinfo {author}
  {\bibfnamefont {E.~T.}\ \bibnamefont {Ritz}}, \bibinfo {author}
  {\bibfnamefont {R.~M.}\ \bibnamefont {Fernandes}}, \bibinfo {author}
  {\bibfnamefont {T.}~\bibnamefont {Birol}}, \bibinfo {author} {\bibfnamefont
  {Z.}~\bibnamefont {Guguchia}},\ and\ \bibinfo {author} {\bibfnamefont
  {H.}~\bibnamefont {Luetkens}},\ }\bibfield  {title} {\bibinfo {title}
  {Time-reversal symmetry broken by charge order in {CsV$_3$Sb$_5$}},\ }\href
  {https://doi.org/10.1103/PhysRevResearch.4.023244} {\bibfield  {journal}
  {\bibinfo  {journal} {Phys. Rev. Research}\ }\textbf {\bibinfo {volume}
  {4}},\ \bibinfo {pages} {023244} (\bibinfo {year} {2022})}\BibitemShut
  {NoStop}%
\bibitem [{\citenamefont {Shan}\ \emph {et~al.}(2022)\citenamefont {Shan},
  \citenamefont {Biswas}, \citenamefont {Ghosh}, \citenamefont {Tula},
  \citenamefont {Hillier}, \citenamefont {Adroja}, \citenamefont {Cottrell},
  \citenamefont {Cao}, \citenamefont {Liu}, \citenamefont {Xu}, \citenamefont
  {Song}, \citenamefont {Yuan},\ and\ \citenamefont {Smidman}}]{Shan2022}%
  \BibitemOpen
  \bibfield  {author} {\bibinfo {author} {\bibfnamefont {Z.}~\bibnamefont
  {Shan}}, \bibinfo {author} {\bibfnamefont {P.~K.}\ \bibnamefont {Biswas}},
  \bibinfo {author} {\bibfnamefont {S.~K.}\ \bibnamefont {Ghosh}}, \bibinfo
  {author} {\bibfnamefont {T.}~\bibnamefont {Tula}}, \bibinfo {author}
  {\bibfnamefont {A.~D.}\ \bibnamefont {Hillier}}, \bibinfo {author}
  {\bibfnamefont {D.}~\bibnamefont {Adroja}}, \bibinfo {author} {\bibfnamefont
  {S.}~\bibnamefont {Cottrell}}, \bibinfo {author} {\bibfnamefont {G.-H.}\
  \bibnamefont {Cao}}, \bibinfo {author} {\bibfnamefont {Y.}~\bibnamefont
  {Liu}}, \bibinfo {author} {\bibfnamefont {X.}~\bibnamefont {Xu}}, \bibinfo
  {author} {\bibfnamefont {Y.}~\bibnamefont {Song}}, \bibinfo {author}
  {\bibfnamefont {H.}~\bibnamefont {Yuan}},\ and\ \bibinfo {author}
  {\bibfnamefont {M.}~\bibnamefont {Smidman}},\ }\bibfield  {title} {\bibinfo
  {title} {Muon spin relaxation study of the layered kagome superconductor
  {CsV$_3$Sb$_5$}},\ }\href {https://doi.org/10.1103/PhysRevResearch.4.033145}
  {\bibfield  {journal} {\bibinfo  {journal} {Phys. Rev. Res.}\ }\textbf
  {\bibinfo {volume} {4}},\ \bibinfo {pages} {033145} (\bibinfo {year}
  {2022})}\BibitemShut {NoStop}%
\bibitem [{\citenamefont {Guguchia}\ \emph {et~al.}(2023)\citenamefont
  {Guguchia}, \citenamefont {Gawryluk}, \citenamefont {Shin}, \citenamefont
  {Hao}, \citenamefont {Mielke~III}, \citenamefont {Das}, \citenamefont
  {Plokhikh}, \citenamefont {Liborio}, \citenamefont {Shenton}, \citenamefont
  {Hu} \emph {et~al.}}]{Guguchia2023}%
  \BibitemOpen
  \bibfield  {author} {\bibinfo {author} {\bibfnamefont {Z.}~\bibnamefont
  {Guguchia}}, \bibinfo {author} {\bibfnamefont {D.}~\bibnamefont {Gawryluk}},
  \bibinfo {author} {\bibfnamefont {S.}~\bibnamefont {Shin}}, \bibinfo {author}
  {\bibfnamefont {Z.}~\bibnamefont {Hao}}, \bibinfo {author} {\bibfnamefont
  {C.}~\bibnamefont {Mielke~III}}, \bibinfo {author} {\bibfnamefont
  {D.}~\bibnamefont {Das}}, \bibinfo {author} {\bibfnamefont {I.}~\bibnamefont
  {Plokhikh}}, \bibinfo {author} {\bibfnamefont {L.}~\bibnamefont {Liborio}},
  \bibinfo {author} {\bibfnamefont {J.~K.}\ \bibnamefont {Shenton}}, \bibinfo
  {author} {\bibfnamefont {Y.}~\bibnamefont {Hu}}, \emph {et~al.},\ }\bibfield
  {title} {\bibinfo {title} {Hidden magnetism uncovered in a charge ordered
  bilayer kagome material scv6sn6},\ }\href@noop {} {\bibfield  {journal}
  {\bibinfo  {journal} {Nature communications}\ }\textbf {\bibinfo {volume}
  {14}},\ \bibinfo {pages} {7796} (\bibinfo {year} {2023})}\BibitemShut
  {NoStop}%
\bibitem [{\citenamefont {Kenney}\ \emph {et~al.}(2021)\citenamefont {Kenney},
  \citenamefont {Ortiz}, \citenamefont {Wang}, \citenamefont {Wilson},\ and\
  \citenamefont {Graf}}]{Kenney2021}%
  \BibitemOpen
  \bibfield  {author} {\bibinfo {author} {\bibfnamefont {E.~M.}\ \bibnamefont
  {Kenney}}, \bibinfo {author} {\bibfnamefont {B.~R.}\ \bibnamefont {Ortiz}},
  \bibinfo {author} {\bibfnamefont {C.}~\bibnamefont {Wang}}, \bibinfo {author}
  {\bibfnamefont {S.~D.}\ \bibnamefont {Wilson}},\ and\ \bibinfo {author}
  {\bibfnamefont {M.~J.}\ \bibnamefont {Graf}},\ }\bibfield  {title} {\bibinfo
  {title} {Absence of local moments in the kagome metal kv3sb5 as determined by
  muon spin spectroscopy},\ }\href {https://doi.org/10.1088/1361-648X/abe8f9}
  {\bibfield  {journal} {\bibinfo  {journal} {Journal of Physics: Condensed
  Matter}\ }\textbf {\bibinfo {volume} {33}},\ \bibinfo {pages} {235801}
  (\bibinfo {year} {2021})}\BibitemShut {NoStop}%
\bibitem [{\citenamefont {Yu}\ \emph {et~al.}(2021{\natexlab{a}})\citenamefont
  {Yu}, \citenamefont {Wu}, \citenamefont {Wang}, \citenamefont {Lei},
  \citenamefont {Zhuo}, \citenamefont {Ying},\ and\ \citenamefont
  {Chen}}]{Yu2021}%
  \BibitemOpen
  \bibfield  {author} {\bibinfo {author} {\bibfnamefont {F.~H.}\ \bibnamefont
  {Yu}}, \bibinfo {author} {\bibfnamefont {T.}~\bibnamefont {Wu}}, \bibinfo
  {author} {\bibfnamefont {Z.~Y.}\ \bibnamefont {Wang}}, \bibinfo {author}
  {\bibfnamefont {B.}~\bibnamefont {Lei}}, \bibinfo {author} {\bibfnamefont
  {W.~Z.}\ \bibnamefont {Zhuo}}, \bibinfo {author} {\bibfnamefont {J.~J.}\
  \bibnamefont {Ying}},\ and\ \bibinfo {author} {\bibfnamefont {X.~H.}\
  \bibnamefont {Chen}},\ }\bibfield  {title} {\bibinfo {title} {Concurrence of
  anomalous hall effect and charge density wave in a superconducting
  topological kagom\'e metal},\ }\href
  {https://doi.org/10.1103/PhysRevB.104.L041103} {\bibfield  {journal}
  {\bibinfo  {journal} {Phys. Rev. B}\ }\textbf {\bibinfo {volume} {104}},\
  \bibinfo {pages} {L041103} (\bibinfo {year}
  {2021}{\natexlab{a}})}\BibitemShut {NoStop}%
\bibitem [{\citenamefont {Yang}\ \emph {et~al.}(2020)\citenamefont {Yang},
  \citenamefont {Wang}, \citenamefont {Ortiz}, \citenamefont {Liu},
  \citenamefont {Gayles}, \citenamefont {Derunova}, \citenamefont
  {Gonzalez-Hernandez}, \citenamefont {Šmejkal}, \citenamefont {Chen},
  \citenamefont {Parkin}, \citenamefont {Wilson}, \citenamefont {Toberer},
  \citenamefont {McQueen},\ and\ \citenamefont {Ali}}]{Yang2020}%
  \BibitemOpen
  \bibfield  {author} {\bibinfo {author} {\bibfnamefont {S.-Y.}\ \bibnamefont
  {Yang}}, \bibinfo {author} {\bibfnamefont {Y.}~\bibnamefont {Wang}}, \bibinfo
  {author} {\bibfnamefont {B.~R.}\ \bibnamefont {Ortiz}}, \bibinfo {author}
  {\bibfnamefont {D.}~\bibnamefont {Liu}}, \bibinfo {author} {\bibfnamefont
  {J.}~\bibnamefont {Gayles}}, \bibinfo {author} {\bibfnamefont
  {E.}~\bibnamefont {Derunova}}, \bibinfo {author} {\bibfnamefont
  {R.}~\bibnamefont {Gonzalez-Hernandez}}, \bibinfo {author} {\bibfnamefont
  {L.}~\bibnamefont {Šmejkal}}, \bibinfo {author} {\bibfnamefont
  {Y.}~\bibnamefont {Chen}}, \bibinfo {author} {\bibfnamefont {S.~S.~P.}\
  \bibnamefont {Parkin}}, \bibinfo {author} {\bibfnamefont {S.~D.}\
  \bibnamefont {Wilson}}, \bibinfo {author} {\bibfnamefont {E.~S.}\
  \bibnamefont {Toberer}}, \bibinfo {author} {\bibfnamefont {T.}~\bibnamefont
  {McQueen}},\ and\ \bibinfo {author} {\bibfnamefont {M.~N.}\ \bibnamefont
  {Ali}},\ }\bibfield  {title} {\bibinfo {title} {Giant, unconventional
  anomalous hall effect in the metallic frustrated magnet candidate,
  {KV$_3$Sb$_5$}},\ }\href {https://doi.org/10.1126/sciadv.abb6003} {\bibfield
  {journal} {\bibinfo  {journal} {Science Advances}\ }\textbf {\bibinfo
  {volume} {6}},\ \bibinfo {pages} {eabb6003} (\bibinfo {year}
  {2020})}\BibitemShut {NoStop}%
\bibitem [{\citenamefont {Yi}\ \emph {et~al.}(2024)\citenamefont {Yi},
  \citenamefont {Feng}, \citenamefont {Kumar}, \citenamefont {Felser},\ and\
  \citenamefont {Shekhar}}]{Yi2024}%
  \BibitemOpen
  \bibfield  {author} {\bibinfo {author} {\bibfnamefont {C.}~\bibnamefont
  {Yi}}, \bibinfo {author} {\bibfnamefont {X.}~\bibnamefont {Feng}}, \bibinfo
  {author} {\bibfnamefont {N.}~\bibnamefont {Kumar}}, \bibinfo {author}
  {\bibfnamefont {C.}~\bibnamefont {Felser}},\ and\ \bibinfo {author}
  {\bibfnamefont {C.}~\bibnamefont {Shekhar}},\ }\bibfield  {title} {\bibinfo
  {title} {Tuning charge density wave of kagome metal {ScV$_6$Sn$_6$}},\ }\href
  {https://doi.org/10.1088/1367-2630/ad4389} {\bibfield  {journal} {\bibinfo
  {journal} {New Journal of Physics}\ }\textbf {\bibinfo {volume} {26}},\
  \bibinfo {pages} {052001} (\bibinfo {year} {2024})}\BibitemShut {NoStop}%
\bibitem [{\citenamefont {Mozaffari}\ \emph {et~al.}(2024)\citenamefont
  {Mozaffari}, \citenamefont {Meier}, \citenamefont {Madhogaria}, \citenamefont
  {Peshcherenko}, \citenamefont {Kang}, \citenamefont {Villanova},
  \citenamefont {Arachchige}, \citenamefont {Zheng}, \citenamefont {Zhu},
  \citenamefont {Chen}, \citenamefont {Jenkins}, \citenamefont {Zhang},
  \citenamefont {Chan}, \citenamefont {Li}, \citenamefont {Yoon}, \citenamefont
  {Zhang},\ and\ \citenamefont {Mandrus}}]{Mozaffari2024}%
  \BibitemOpen
  \bibfield  {author} {\bibinfo {author} {\bibfnamefont {S.}~\bibnamefont
  {Mozaffari}}, \bibinfo {author} {\bibfnamefont {W.~R.}\ \bibnamefont
  {Meier}}, \bibinfo {author} {\bibfnamefont {R.~P.}\ \bibnamefont
  {Madhogaria}}, \bibinfo {author} {\bibfnamefont {N.}~\bibnamefont
  {Peshcherenko}}, \bibinfo {author} {\bibfnamefont {S.-H.}\ \bibnamefont
  {Kang}}, \bibinfo {author} {\bibfnamefont {J.~W.}\ \bibnamefont {Villanova}},
  \bibinfo {author} {\bibfnamefont {H.~W.~S.}\ \bibnamefont {Arachchige}},
  \bibinfo {author} {\bibfnamefont {G.}~\bibnamefont {Zheng}}, \bibinfo
  {author} {\bibfnamefont {Y.}~\bibnamefont {Zhu}}, \bibinfo {author}
  {\bibfnamefont {K.-W.}\ \bibnamefont {Chen}}, \bibinfo {author}
  {\bibfnamefont {K.}~\bibnamefont {Jenkins}}, \bibinfo {author} {\bibfnamefont
  {D.}~\bibnamefont {Zhang}}, \bibinfo {author} {\bibfnamefont
  {A.}~\bibnamefont {Chan}}, \bibinfo {author} {\bibfnamefont {L.}~\bibnamefont
  {Li}}, \bibinfo {author} {\bibfnamefont {M.}~\bibnamefont {Yoon}}, \bibinfo
  {author} {\bibfnamefont {Y.}~\bibnamefont {Zhang}},\ and\ \bibinfo {author}
  {\bibfnamefont {D.~G.}\ \bibnamefont {Mandrus}},\ }\bibfield  {title}
  {\bibinfo {title} {Universal sublinear resistivity in vanadium kagome
  materials hosting charge density waves},\ }\href
  {https://doi.org/10.1103/PhysRevB.110.035135} {\bibfield  {journal} {\bibinfo
   {journal} {Phys. Rev. B}\ }\textbf {\bibinfo {volume} {110}},\ \bibinfo
  {pages} {035135} (\bibinfo {year} {2024})}\BibitemShut {NoStop}%
\bibitem [{\citenamefont {Liu}\ \emph {et~al.}(2025{\natexlab{a}})\citenamefont
  {Liu}, \citenamefont {Roppongi}, \citenamefont {Kimata}, \citenamefont
  {Ishihara}, \citenamefont {Grasset}, \citenamefont {Konczykowski},
  \citenamefont {Ortiz}, \citenamefont {Wilson}, \citenamefont {Yoshimi},
  \citenamefont {Shibauchi},\ and\ \citenamefont {Hashimoto}}]{Liu2025b}%
  \BibitemOpen
  \bibfield  {author} {\bibinfo {author} {\bibfnamefont {S.}~\bibnamefont
  {Liu}}, \bibinfo {author} {\bibfnamefont {M.}~\bibnamefont {Roppongi}},
  \bibinfo {author} {\bibfnamefont {M.}~\bibnamefont {Kimata}}, \bibinfo
  {author} {\bibfnamefont {K.}~\bibnamefont {Ishihara}}, \bibinfo {author}
  {\bibfnamefont {R.}~\bibnamefont {Grasset}}, \bibinfo {author} {\bibfnamefont
  {M.}~\bibnamefont {Konczykowski}}, \bibinfo {author} {\bibfnamefont {B.~R.}\
  \bibnamefont {Ortiz}}, \bibinfo {author} {\bibfnamefont {S.~D.}\ \bibnamefont
  {Wilson}}, \bibinfo {author} {\bibfnamefont {K.}~\bibnamefont {Yoshimi}},
  \bibinfo {author} {\bibfnamefont {T.}~\bibnamefont {Shibauchi}},\ and\
  \bibinfo {author} {\bibfnamefont {K.}~\bibnamefont {Hashimoto}},\ }\bibfield
  {title} {\bibinfo {title} {Impact of tiny fermi pockets with extremely high
  mobility on the hall anomaly in the kagome metal {CsV$_3$Sb$_5$}},\ }\href
  {https://doi.org/10.1103/d4dw-2v6k} {\bibfield  {journal} {\bibinfo
  {journal} {Phys. Rev. Lett.}\ }\textbf {\bibinfo {volume} {135}},\ \bibinfo
  {pages} {056502} (\bibinfo {year} {2025}{\natexlab{a}})}\BibitemShut
  {NoStop}%
\bibitem [{\citenamefont {Koshelev}\ \emph {et~al.}(2024)\citenamefont
  {Koshelev}, \citenamefont {Chapai}, \citenamefont {Chung}, \citenamefont
  {Mitchell},\ and\ \citenamefont {Welp}}]{Koshelev2024}%
  \BibitemOpen
  \bibfield  {author} {\bibinfo {author} {\bibfnamefont {A.~E.}\ \bibnamefont
  {Koshelev}}, \bibinfo {author} {\bibfnamefont {R.}~\bibnamefont {Chapai}},
  \bibinfo {author} {\bibfnamefont {D.~Y.}\ \bibnamefont {Chung}}, \bibinfo
  {author} {\bibfnamefont {J.~F.}\ \bibnamefont {Mitchell}},\ and\ \bibinfo
  {author} {\bibfnamefont {U.}~\bibnamefont {Welp}},\ }\bibfield  {title}
  {\bibinfo {title} {Origin of anomalous magnetotransport in kagome
  superconductors {AV$_3$Sb$_5$ ($A=\mathrm{K}$,Rb,Cs)}},\ }\href
  {https://doi.org/10.1103/PhysRevB.110.024512} {\bibfield  {journal} {\bibinfo
   {journal} {Phys. Rev. B}\ }\textbf {\bibinfo {volume} {110}},\ \bibinfo
  {pages} {024512} (\bibinfo {year} {2024})}\BibitemShut {NoStop}%
\bibitem [{\citenamefont {Wu}\ \emph {et~al.}(2022)\citenamefont {Wu},
  \citenamefont {Wang}, \citenamefont {Liu}, \citenamefont {Li}, \citenamefont
  {Xu}, \citenamefont {Yin}, \citenamefont {Gong}, \citenamefont {Tu},
  \citenamefont {Lei}, \citenamefont {Dong},\ and\ \citenamefont
  {Wang}}]{Wu2021}%
  \BibitemOpen
  \bibfield  {author} {\bibinfo {author} {\bibfnamefont {Q.}~\bibnamefont
  {Wu}}, \bibinfo {author} {\bibfnamefont {Z.~X.}\ \bibnamefont {Wang}},
  \bibinfo {author} {\bibfnamefont {Q.~M.}\ \bibnamefont {Liu}}, \bibinfo
  {author} {\bibfnamefont {R.~S.}\ \bibnamefont {Li}}, \bibinfo {author}
  {\bibfnamefont {S.~X.}\ \bibnamefont {Xu}}, \bibinfo {author} {\bibfnamefont
  {Q.~W.}\ \bibnamefont {Yin}}, \bibinfo {author} {\bibfnamefont {C.~S.}\
  \bibnamefont {Gong}}, \bibinfo {author} {\bibfnamefont {Z.~J.}\ \bibnamefont
  {Tu}}, \bibinfo {author} {\bibfnamefont {H.~C.}\ \bibnamefont {Lei}},
  \bibinfo {author} {\bibfnamefont {T.}~\bibnamefont {Dong}},\ and\ \bibinfo
  {author} {\bibfnamefont {N.~L.}\ \bibnamefont {Wang}},\ }\bibfield  {title}
  {\bibinfo {title} {Simultaneous formation of two-fold rotation symmetry with
  charge order in the kagome superconductor
  {${\mathrm{CsV}}_{3}{\mathrm{Sb}}_{5}$} by optical polarization rotation
  measurement},\ }\href {https://doi.org/10.1103/PhysRevB.106.205109}
  {\bibfield  {journal} {\bibinfo  {journal} {Phys. Rev. B}\ }\textbf {\bibinfo
  {volume} {106}},\ \bibinfo {pages} {205109} (\bibinfo {year}
  {2022})}\BibitemShut {NoStop}%
\bibitem [{\citenamefont {Xu}\ \emph {et~al.}(2022)\citenamefont {Xu},
  \citenamefont {Ni}, \citenamefont {Liu}, \citenamefont {Ortiz}, \citenamefont
  {Deng}, \citenamefont {Wilson}, \citenamefont {Yan}, \citenamefont
  {Balents},\ and\ \citenamefont {Wu}}]{Xu2022}%
  \BibitemOpen
  \bibfield  {author} {\bibinfo {author} {\bibfnamefont {Y.}~\bibnamefont
  {Xu}}, \bibinfo {author} {\bibfnamefont {Z.}~\bibnamefont {Ni}}, \bibinfo
  {author} {\bibfnamefont {Y.}~\bibnamefont {Liu}}, \bibinfo {author}
  {\bibfnamefont {B.~R.}\ \bibnamefont {Ortiz}}, \bibinfo {author}
  {\bibfnamefont {Q.}~\bibnamefont {Deng}}, \bibinfo {author} {\bibfnamefont
  {S.~D.}\ \bibnamefont {Wilson}}, \bibinfo {author} {\bibfnamefont
  {B.}~\bibnamefont {Yan}}, \bibinfo {author} {\bibfnamefont {L.}~\bibnamefont
  {Balents}},\ and\ \bibinfo {author} {\bibfnamefont {L.}~\bibnamefont {Wu}},\
  }\bibfield  {title} {\bibinfo {title} {Three-state nematicity and
  magneto-optical kerr effect in the charge density waves in kagome
  superconductors},\ }\href {https://doi.org/10.1038/s41567-022-01805-7}
  {\bibfield  {journal} {\bibinfo  {journal} {Nature Physics}\ }\textbf
  {\bibinfo {volume} {18}},\ \bibinfo {pages} {1470} (\bibinfo {year}
  {2022})}\BibitemShut {NoStop}%
\bibitem [{\citenamefont {Oppeneer}(2001)}]{Oppeneer2001}%
  \BibitemOpen
  \bibfield  {author} {\bibinfo {author} {\bibfnamefont {P.}~\bibnamefont
  {Oppeneer}},\ }\bibfield  {title} {\bibinfo {title} {Magneto-optical kerr
  spectra},\ }\href@noop {} {\bibfield  {journal} {\bibinfo  {journal}
  {Handbook of Magnetic Materials}\ }\textbf {\bibinfo {volume} {13}},\
  \bibinfo {pages} {229} (\bibinfo {year} {2001})}\BibitemShut {NoStop}%
\bibitem [{\citenamefont {Uykur}\ \emph {et~al.}(2021)\citenamefont {Uykur},
  \citenamefont {Ortiz}, \citenamefont {Iakutkina}, \citenamefont {Wenzel},
  \citenamefont {Wilson}, \citenamefont {Dressel},\ and\ \citenamefont
  {Tsirlin}}]{Uykur2021}%
  \BibitemOpen
  \bibfield  {author} {\bibinfo {author} {\bibfnamefont {E.}~\bibnamefont
  {Uykur}}, \bibinfo {author} {\bibfnamefont {B.~R.}\ \bibnamefont {Ortiz}},
  \bibinfo {author} {\bibfnamefont {O.}~\bibnamefont {Iakutkina}}, \bibinfo
  {author} {\bibfnamefont {M.}~\bibnamefont {Wenzel}}, \bibinfo {author}
  {\bibfnamefont {S.~D.}\ \bibnamefont {Wilson}}, \bibinfo {author}
  {\bibfnamefont {M.}~\bibnamefont {Dressel}},\ and\ \bibinfo {author}
  {\bibfnamefont {A.~A.}\ \bibnamefont {Tsirlin}},\ }\bibfield  {title}
  {\bibinfo {title} {Low-energy optical properties of the nonmagnetic kagom\'e
  metal {CsV$_3$Sb$_5$}},\ }\href {https://doi.org/10.1103/PhysRevB.104.045130}
  {\bibfield  {journal} {\bibinfo  {journal} {Phys. Rev. B}\ }\textbf {\bibinfo
  {volume} {104}},\ \bibinfo {pages} {045130} (\bibinfo {year}
  {2021})}\BibitemShut {NoStop}%
\bibitem [{\citenamefont {Zhou}\ \emph {et~al.}(2021)\citenamefont {Zhou},
  \citenamefont {Li}, \citenamefont {Fan}, \citenamefont {Hao}, \citenamefont
  {Dai}, \citenamefont {Wang}, \citenamefont {Yao},\ and\ \citenamefont
  {Wen}}]{Zhou2021}%
  \BibitemOpen
  \bibfield  {author} {\bibinfo {author} {\bibfnamefont {X.}~\bibnamefont
  {Zhou}}, \bibinfo {author} {\bibfnamefont {Y.}~\bibnamefont {Li}}, \bibinfo
  {author} {\bibfnamefont {X.}~\bibnamefont {Fan}}, \bibinfo {author}
  {\bibfnamefont {J.}~\bibnamefont {Hao}}, \bibinfo {author} {\bibfnamefont
  {Y.}~\bibnamefont {Dai}}, \bibinfo {author} {\bibfnamefont {Z.}~\bibnamefont
  {Wang}}, \bibinfo {author} {\bibfnamefont {Y.}~\bibnamefont {Yao}},\ and\
  \bibinfo {author} {\bibfnamefont {H.-H.}\ \bibnamefont {Wen}},\ }\bibfield
  {title} {\bibinfo {title} {Origin of charge density wave in the kagome metal
  ${\mathrm{csv}}_{3}{\mathrm{sb}}_{5}$ as revealed by optical spectroscopy},\
  }\href {https://doi.org/10.1103/PhysRevB.104.L041101} {\bibfield  {journal}
  {\bibinfo  {journal} {Phys. Rev. B}\ }\textbf {\bibinfo {volume} {104}},\
  \bibinfo {pages} {L041101} (\bibinfo {year} {2021})}\BibitemShut {NoStop}%
\bibitem [{\citenamefont {Saykin}\ \emph {et~al.}(2023)\citenamefont {Saykin},
  \citenamefont {Farhang}, \citenamefont {Kountz}, \citenamefont {Chen},
  \citenamefont {Ortiz}, \citenamefont {Shekhar}, \citenamefont {Felser},
  \citenamefont {Wilson}, \citenamefont {Thomale}, \citenamefont {Xia},\ and\
  \citenamefont {Kapitulnik}}]{Saykin2023}%
  \BibitemOpen
  \bibfield  {author} {\bibinfo {author} {\bibfnamefont {D.~R.}\ \bibnamefont
  {Saykin}}, \bibinfo {author} {\bibfnamefont {C.}~\bibnamefont {Farhang}},
  \bibinfo {author} {\bibfnamefont {E.~D.}\ \bibnamefont {Kountz}}, \bibinfo
  {author} {\bibfnamefont {D.}~\bibnamefont {Chen}}, \bibinfo {author}
  {\bibfnamefont {B.~R.}\ \bibnamefont {Ortiz}}, \bibinfo {author}
  {\bibfnamefont {C.}~\bibnamefont {Shekhar}}, \bibinfo {author} {\bibfnamefont
  {C.}~\bibnamefont {Felser}}, \bibinfo {author} {\bibfnamefont {S.~D.}\
  \bibnamefont {Wilson}}, \bibinfo {author} {\bibfnamefont {R.}~\bibnamefont
  {Thomale}}, \bibinfo {author} {\bibfnamefont {J.}~\bibnamefont {Xia}},\ and\
  \bibinfo {author} {\bibfnamefont {A.}~\bibnamefont {Kapitulnik}},\ }\bibfield
   {title} {\bibinfo {title} {High resolution polar kerr effect studies of
  {CsV$_3$Sb$_5$}: Tests for time-reversal symmetry breaking below the
  charge-order transition},\ }\href
  {https://doi.org/10.1103/PhysRevLett.131.016901} {\bibfield  {journal}
  {\bibinfo  {journal} {Phys. Rev. Lett.}\ }\textbf {\bibinfo {volume} {131}},\
  \bibinfo {pages} {016901} (\bibinfo {year} {2023})}\BibitemShut {NoStop}%
\bibitem [{\citenamefont {Farhang}\ \emph {et~al.}(2023)\citenamefont
  {Farhang}, \citenamefont {Wang}, \citenamefont {Ortiz}, \citenamefont
  {Wilson},\ and\ \citenamefont {Xia}}]{Farhang2023}%
  \BibitemOpen
  \bibfield  {author} {\bibinfo {author} {\bibfnamefont {C.}~\bibnamefont
  {Farhang}}, \bibinfo {author} {\bibfnamefont {J.}~\bibnamefont {Wang}},
  \bibinfo {author} {\bibfnamefont {B.~R.}\ \bibnamefont {Ortiz}}, \bibinfo
  {author} {\bibfnamefont {S.~D.}\ \bibnamefont {Wilson}},\ and\ \bibinfo
  {author} {\bibfnamefont {J.}~\bibnamefont {Xia}},\ }\bibfield  {title}
  {\bibinfo {title} {Unconventional specular optical rotation in the charge
  ordered state of kagome metal {CsV$_3$Sb$_5$}},\ }\href
  {https://doi.org/10.1038/s41467-023-41080-5} {\bibfield  {journal} {\bibinfo
  {journal} {Nature Communications}\ }\textbf {\bibinfo {volume} {14}},\
  \bibinfo {pages} {5326} (\bibinfo {year} {2023})}\BibitemShut {NoStop}%
\bibitem [{\citenamefont {Liu}\ \emph {et~al.}(2025{\natexlab{b}})\citenamefont
  {Liu}, \citenamefont {Huang}, \citenamefont {Zhou},\ and\ \citenamefont
  {Huang}}]{Liu2025a}%
  \BibitemOpen
  \bibfield  {author} {\bibinfo {author} {\bibfnamefont {H.-T.}\ \bibnamefont
  {Liu}}, \bibinfo {author} {\bibfnamefont {J.}~\bibnamefont {Huang}}, \bibinfo
  {author} {\bibfnamefont {T.}~\bibnamefont {Zhou}},\ and\ \bibinfo {author}
  {\bibfnamefont {W.}~\bibnamefont {Huang}},\ }\bibfield  {title} {\bibinfo
  {title} {Constraints on the orbital flux phase in
  {$A{\mathrm{V}}_{3}{\mathrm{Sb}}_{5}$} from the polar kerr effect},\ }\href
  {https://doi.org/10.1103/PhysRevB.111.L041109} {\bibfield  {journal}
  {\bibinfo  {journal} {Phys. Rev. B}\ }\textbf {\bibinfo {volume} {111}},\
  \bibinfo {pages} {L041109} (\bibinfo {year}
  {2025}{\natexlab{b}})}\BibitemShut {NoStop}%
\bibitem [{\citenamefont {Xia}\ \emph {et~al.}(2006{\natexlab{a}})\citenamefont
  {Xia}, \citenamefont {Beyersdorf}, \citenamefont {Fejer},\ and\ \citenamefont
  {Kapitulnik}}]{Xia2006a}%
  \BibitemOpen
  \bibfield  {author} {\bibinfo {author} {\bibfnamefont {J.}~\bibnamefont
  {Xia}}, \bibinfo {author} {\bibfnamefont {P.~T.}\ \bibnamefont {Beyersdorf}},
  \bibinfo {author} {\bibfnamefont {M.~M.}\ \bibnamefont {Fejer}},\ and\
  \bibinfo {author} {\bibfnamefont {A.}~\bibnamefont {Kapitulnik}},\ }\bibfield
   {title} {\bibinfo {title} {Modified sagnac interferometer for
  high-sensitivity magneto-optic measurements at cryogenic temperatures},\
  }\href {https://doi.org/10.1063/1.2336620} {\bibfield  {journal} {\bibinfo
  {journal} {Applied Physics Letters}\ }\textbf {\bibinfo {volume} {89}},\
  \bibinfo {pages} {062508} (\bibinfo {year} {2006}{\natexlab{a}})}\BibitemShut
  {NoStop}%
\bibitem [{\citenamefont {Sagnac}(1913)}]{Sagnac1913}%
  \BibitemOpen
  \bibfield  {author} {\bibinfo {author} {\bibfnamefont {G.}~\bibnamefont
  {Sagnac}},\ }\bibfield  {title} {\bibinfo {title} {L'\'ether lumineux
  d\'emontr\'e par l'effet du vent relatif d'éther dans un interf\'erom\`etre
  en rotation uniforme},\ }\href@noop {} {\bibfield  {journal} {\bibinfo
  {journal} {Comptes Rendus}\ }\textbf {\bibinfo {volume} {157}},\ \bibinfo
  {pages} {708} (\bibinfo {year} {1913})}\BibitemShut {NoStop}%
\bibitem [{sup()}]{supplement}%
  \BibitemOpen
  \href@noop {} {}\bibinfo {note} {See Supplementary Material at [URL will be
  inserted by publisher] for details on the device characterization and data
  analysis}\BibitemShut {NoStop}%
\bibitem [{\citenamefont {Karapetyan}\ \emph {et~al.}(2012)\citenamefont
  {Karapetyan}, \citenamefont {H\"ucker}, \citenamefont {Gu}, \citenamefont
  {Tranquada}, \citenamefont {Fejer}, \citenamefont {Xia},\ and\ \citenamefont
  {Kapitulnik}}]{Karapetyan2012}%
  \BibitemOpen
  \bibfield  {author} {\bibinfo {author} {\bibfnamefont {H.}~\bibnamefont
  {Karapetyan}}, \bibinfo {author} {\bibfnamefont {M.}~\bibnamefont
  {H\"ucker}}, \bibinfo {author} {\bibfnamefont {G.~D.}\ \bibnamefont {Gu}},
  \bibinfo {author} {\bibfnamefont {J.~M.}\ \bibnamefont {Tranquada}}, \bibinfo
  {author} {\bibfnamefont {M.~M.}\ \bibnamefont {Fejer}}, \bibinfo {author}
  {\bibfnamefont {J.}~\bibnamefont {Xia}},\ and\ \bibinfo {author}
  {\bibfnamefont {A.}~\bibnamefont {Kapitulnik}},\ }\bibfield  {title}
  {\bibinfo {title} {Magneto-optical measurements of a cascade of transitions
  in superconducting
  {${\mathrm{La}}_{1.875}{\mathrm{Ba}}_{0.125}{\mathrm{CuO}}_{4}$ Single
  Crystals}},\ }\href {https://doi.org/10.1103/PhysRevLett.109.147001}
  {\bibfield  {journal} {\bibinfo  {journal} {Phys. Rev. Lett.}\ }\textbf
  {\bibinfo {volume} {109}},\ \bibinfo {pages} {147001} (\bibinfo {year}
  {2012})}\BibitemShut {NoStop}%
\bibitem [{\citenamefont {Kapitulnik}\ \emph {et~al.}(2009)\citenamefont
  {Kapitulnik}, \citenamefont {Xia}, \citenamefont {Schemm},\ and\
  \citenamefont {Palevski}}]{Kapitulnik2009}%
  \BibitemOpen
  \bibfield  {author} {\bibinfo {author} {\bibfnamefont {A.}~\bibnamefont
  {Kapitulnik}}, \bibinfo {author} {\bibfnamefont {J.}~\bibnamefont {Xia}},
  \bibinfo {author} {\bibfnamefont {E.}~\bibnamefont {Schemm}},\ and\ \bibinfo
  {author} {\bibfnamefont {A.}~\bibnamefont {Palevski}},\ }\bibfield  {title}
  {\bibinfo {title} {Polar kerr effect as probe for time-reversal symmetry
  breaking in unconventional superconductors},\ }\href
  {https://doi.org/10.1088/1367-2630/11/5/055060} {\bibfield  {journal}
  {\bibinfo  {journal} {New Journal of Physics}\ }\textbf {\bibinfo {volume}
  {11}},\ \bibinfo {pages} {055060} (\bibinfo {year} {2009})}\BibitemShut
  {NoStop}%
\bibitem [{\citenamefont {Xia}\ \emph {et~al.}(2006{\natexlab{b}})\citenamefont
  {Xia}, \citenamefont {Maeno}, \citenamefont {Beyersdorf}, \citenamefont
  {Fejer},\ and\ \citenamefont {Kapitulnik}}]{Xia2006b}%
  \BibitemOpen
  \bibfield  {author} {\bibinfo {author} {\bibfnamefont {J.}~\bibnamefont
  {Xia}}, \bibinfo {author} {\bibfnamefont {Y.}~\bibnamefont {Maeno}}, \bibinfo
  {author} {\bibfnamefont {P.~T.}\ \bibnamefont {Beyersdorf}}, \bibinfo
  {author} {\bibfnamefont {M.~M.}\ \bibnamefont {Fejer}},\ and\ \bibinfo
  {author} {\bibfnamefont {A.}~\bibnamefont {Kapitulnik}},\ }\bibfield  {title}
  {\bibinfo {title} {High resolution polar kerr effect measurements of
  ${\mathrm{sr}}_{2}{\mathrm{ruo}}_{4}$: Evidence for broken time-reversal
  symmetry in the superconducting state},\ }\href
  {https://doi.org/10.1103/PhysRevLett.97.167002} {\bibfield  {journal}
  {\bibinfo  {journal} {Phys. Rev. Lett.}\ }\textbf {\bibinfo {volume} {97}},\
  \bibinfo {pages} {167002} (\bibinfo {year} {2006}{\natexlab{b}})}\BibitemShut
  {NoStop}%
\bibitem [{\citenamefont {Schemm}\ \emph {et~al.}(2014)\citenamefont {Schemm},
  \citenamefont {Gannon}, \citenamefont {Wishne}, \citenamefont {Halperin},\
  and\ \citenamefont {Kapitulnik}}]{Schemm2014}%
  \BibitemOpen
  \bibfield  {author} {\bibinfo {author} {\bibfnamefont {E.~R.}\ \bibnamefont
  {Schemm}}, \bibinfo {author} {\bibfnamefont {W.~J.}\ \bibnamefont {Gannon}},
  \bibinfo {author} {\bibfnamefont {C.~M.}\ \bibnamefont {Wishne}}, \bibinfo
  {author} {\bibfnamefont {W.~P.}\ \bibnamefont {Halperin}},\ and\ \bibinfo
  {author} {\bibfnamefont {A.}~\bibnamefont {Kapitulnik}},\ }\bibfield  {title}
  {\bibinfo {title} {Observation of broken time-reversal symmetry in the
  heavy-fermion superconductor {UPt$_3$}},\ }\href
  {https://doi.org/10.1126/science.1248552} {\bibfield  {journal} {\bibinfo
  {journal} {Science}\ }\textbf {\bibinfo {volume} {345}},\ \bibinfo {pages}
  {190} (\bibinfo {year} {2014})}\BibitemShut {NoStop}%
\bibitem [{\citenamefont {Xia}\ \emph {et~al.}(2009{\natexlab{a}})\citenamefont
  {Xia}, \citenamefont {Siemons}, \citenamefont {Koster}, \citenamefont
  {Beasley},\ and\ \citenamefont {Kapitulnik}}]{Xia2009b}%
  \BibitemOpen
  \bibfield  {author} {\bibinfo {author} {\bibfnamefont {J.}~\bibnamefont
  {Xia}}, \bibinfo {author} {\bibfnamefont {W.}~\bibnamefont {Siemons}},
  \bibinfo {author} {\bibfnamefont {G.}~\bibnamefont {Koster}}, \bibinfo
  {author} {\bibfnamefont {M.~R.}\ \bibnamefont {Beasley}},\ and\ \bibinfo
  {author} {\bibfnamefont {A.}~\bibnamefont {Kapitulnik}},\ }\bibfield  {title}
  {\bibinfo {title} {Critical thickness for itinerant ferromagnetism in
  ultrathin films of {SrRuO$_3$}},\ }\href
  {https://doi.org/10.1103/PhysRevB.79.140407} {\bibfield  {journal} {\bibinfo
  {journal} {Phys. Rev. B}\ }\textbf {\bibinfo {volume} {79}},\ \bibinfo
  {pages} {140407(R)} (\bibinfo {year} {2009}{\natexlab{a}})}\BibitemShut
  {NoStop}%
\bibitem [{\citenamefont {Xia}\ \emph {et~al.}(2009{\natexlab{b}})\citenamefont
  {Xia}, \citenamefont {Shelukhin}, \citenamefont {Karpovski}, \citenamefont
  {Kapitulnik},\ and\ \citenamefont {Palevski}}]{Xia2009a}%
  \BibitemOpen
  \bibfield  {author} {\bibinfo {author} {\bibfnamefont {J.}~\bibnamefont
  {Xia}}, \bibinfo {author} {\bibfnamefont {V.}~\bibnamefont {Shelukhin}},
  \bibinfo {author} {\bibfnamefont {M.}~\bibnamefont {Karpovski}}, \bibinfo
  {author} {\bibfnamefont {A.}~\bibnamefont {Kapitulnik}},\ and\ \bibinfo
  {author} {\bibfnamefont {A.}~\bibnamefont {Palevski}},\ }\bibfield  {title}
  {\bibinfo {title} {Inverse proximity effect in superconductor-ferromagnet
  bilayer structures},\ }\href {https://doi.org/10.1103/PhysRevLett.102.087004}
  {\bibfield  {journal} {\bibinfo  {journal} {Phys. Rev. Lett.}\ }\textbf
  {\bibinfo {volume} {102}},\ \bibinfo {pages} {087004} (\bibinfo {year}
  {2009}{\natexlab{b}})}\BibitemShut {NoStop}%
\bibitem [{\citenamefont {Sarkar}\ \emph {et~al.}(2020)\citenamefont {Sarkar},
  \citenamefont {Wei}, \citenamefont {Zhang}, \citenamefont {Poniatowski},
  \citenamefont {Mandal}, \citenamefont {Kapitulnik},\ and\ \citenamefont
  {Greene}}]{Sarkar2020}%
  \BibitemOpen
  \bibfield  {author} {\bibinfo {author} {\bibfnamefont {T.}~\bibnamefont
  {Sarkar}}, \bibinfo {author} {\bibfnamefont {D.~S.}\ \bibnamefont {Wei}},
  \bibinfo {author} {\bibfnamefont {J.}~\bibnamefont {Zhang}}, \bibinfo
  {author} {\bibfnamefont {N.~R.}\ \bibnamefont {Poniatowski}}, \bibinfo
  {author} {\bibfnamefont {P.~R.}\ \bibnamefont {Mandal}}, \bibinfo {author}
  {\bibfnamefont {A.}~\bibnamefont {Kapitulnik}},\ and\ \bibinfo {author}
  {\bibfnamefont {R.~L.}\ \bibnamefont {Greene}},\ }\bibfield  {title}
  {\bibinfo {title} {Ferromagnetic order beyond the superconducting dome in a
  cuprate superconductor},\ }\href {https://doi.org/10.1126/science.aax1581}
  {\bibfield  {journal} {\bibinfo  {journal} {Science}\ }\textbf {\bibinfo
  {volume} {368}},\ \bibinfo {pages} {532} (\bibinfo {year} {2020})},\ \Eprint
  {https://arxiv.org/abs/https://www.science.org/doi/pdf/10.1126/science.aax1581}
  {https://www.science.org/doi/pdf/10.1126/science.aax1581} \BibitemShut
  {NoStop}%
\bibitem [{\citenamefont {Hu}\ \emph {et~al.}(2023)\citenamefont {Hu},
  \citenamefont {Pi}, \citenamefont {Xu}, \citenamefont {Yue}, \citenamefont
  {Wu}, \citenamefont {Liu}, \citenamefont {Zhang}, \citenamefont {Li},
  \citenamefont {Zhou}, \citenamefont {Yuan}, \citenamefont {Wu}, \citenamefont
  {Dong}, \citenamefont {Weng},\ and\ \citenamefont {Wang}}]{Hu2023}%
  \BibitemOpen
  \bibfield  {author} {\bibinfo {author} {\bibfnamefont {T.}~\bibnamefont
  {Hu}}, \bibinfo {author} {\bibfnamefont {H.}~\bibnamefont {Pi}}, \bibinfo
  {author} {\bibfnamefont {S.}~\bibnamefont {Xu}}, \bibinfo {author}
  {\bibfnamefont {L.}~\bibnamefont {Yue}}, \bibinfo {author} {\bibfnamefont
  {Q.}~\bibnamefont {Wu}}, \bibinfo {author} {\bibfnamefont {Q.}~\bibnamefont
  {Liu}}, \bibinfo {author} {\bibfnamefont {S.}~\bibnamefont {Zhang}}, \bibinfo
  {author} {\bibfnamefont {R.}~\bibnamefont {Li}}, \bibinfo {author}
  {\bibfnamefont {X.}~\bibnamefont {Zhou}}, \bibinfo {author} {\bibfnamefont
  {J.}~\bibnamefont {Yuan}}, \bibinfo {author} {\bibfnamefont {D.}~\bibnamefont
  {Wu}}, \bibinfo {author} {\bibfnamefont {T.}~\bibnamefont {Dong}}, \bibinfo
  {author} {\bibfnamefont {H.}~\bibnamefont {Weng}},\ and\ \bibinfo {author}
  {\bibfnamefont {N.}~\bibnamefont {Wang}},\ }\bibfield  {title} {\bibinfo
  {title} {Optical spectroscopy and band structure calculations of the
  structural phase transition in the vanadium-based kagome metal
  {${\mathrm{ScV}}_{6}{\mathrm{Sn}}_{6}$}},\ }\href
  {https://doi.org/10.1103/PhysRevB.107.165119} {\bibfield  {journal} {\bibinfo
   {journal} {Phys. Rev. B}\ }\textbf {\bibinfo {volume} {107}},\ \bibinfo
  {pages} {165119} (\bibinfo {year} {2023})}\BibitemShut {NoStop}%
\bibitem [{\citenamefont {Xia}\ \emph {et~al.}(2006{\natexlab{c}})\citenamefont
  {Xia}, \citenamefont {Beyersdorf}, \citenamefont {Fejer},\ and\ \citenamefont
  {Kapitulnik}}]{Xia2006}%
  \BibitemOpen
  \bibfield  {author} {\bibinfo {author} {\bibfnamefont {J.}~\bibnamefont
  {Xia}}, \bibinfo {author} {\bibfnamefont {P.~T.}\ \bibnamefont {Beyersdorf}},
  \bibinfo {author} {\bibfnamefont {M.~M.}\ \bibnamefont {Fejer}},\ and\
  \bibinfo {author} {\bibfnamefont {A.}~\bibnamefont {Kapitulnik}},\ }\bibfield
   {title} {\bibinfo {title} {Modified sagnac interferometer for
  high-sensitivity magneto-optic measurements at cryogenic temperatures},\
  }\href {https://doi.org/10.1063/1.2336620} {\bibfield  {journal} {\bibinfo
  {journal} {Applied Physics Letters}\ }\textbf {\bibinfo {volume} {89}},\
  \bibinfo {pages} {062508} (\bibinfo {year} {2006}{\natexlab{c}})}\BibitemShut
  {NoStop}%
\bibitem [{\citenamefont {Pershan}(1967)}]{Pershan1967}%
  \BibitemOpen
  \bibfield  {author} {\bibinfo {author} {\bibfnamefont {P.~S.}\ \bibnamefont
  {Pershan}},\ }\bibfield  {title} {\bibinfo {title} {Magneto‐optical
  effects},\ }\href {https://doi.org/10.1063/1.1709678} {\bibfield  {journal}
  {\bibinfo  {journal} {Journal of Applied Physics}\ }\textbf {\bibinfo
  {volume} {38}},\ \bibinfo {pages} {1482} (\bibinfo {year}
  {1967})}\BibitemShut {NoStop}%
\bibitem [{\citenamefont {Stern}\ \emph {et~al.}(1964)\citenamefont {Stern},
  \citenamefont {McGroddy},\ and\ \citenamefont {Harte}}]{Stern1964}%
  \BibitemOpen
  \bibfield  {author} {\bibinfo {author} {\bibfnamefont {E.~A.}\ \bibnamefont
  {Stern}}, \bibinfo {author} {\bibfnamefont {J.~C.}\ \bibnamefont
  {McGroddy}},\ and\ \bibinfo {author} {\bibfnamefont {W.~E.}\ \bibnamefont
  {Harte}},\ }\bibfield  {title} {\bibinfo {title} {Polar reflection faraday
  effect in metals},\ }\href {https://doi.org/10.1103/PhysRev.135.A1306}
  {\bibfield  {journal} {\bibinfo  {journal} {Phys. Rev.}\ }\textbf {\bibinfo
  {volume} {135}},\ \bibinfo {pages} {A1306} (\bibinfo {year}
  {1964})}\BibitemShut {NoStop}%
\bibitem [{\citenamefont {McGroddy}\ \emph {et~al.}(1965)\citenamefont
  {McGroddy}, \citenamefont {McAlister},\ and\ \citenamefont
  {Stern}}]{McGroddy1965}%
  \BibitemOpen
  \bibfield  {author} {\bibinfo {author} {\bibfnamefont {J.~C.}\ \bibnamefont
  {McGroddy}}, \bibinfo {author} {\bibfnamefont {A.~J.}\ \bibnamefont
  {McAlister}},\ and\ \bibinfo {author} {\bibfnamefont {E.~A.}\ \bibnamefont
  {Stern}},\ }\bibfield  {title} {\bibinfo {title} {Polar reflection faraday
  effect in silver and gold},\ }\href
  {https://doi.org/10.1103/PhysRev.139.A1844} {\bibfield  {journal} {\bibinfo
  {journal} {Phys. Rev.}\ }\textbf {\bibinfo {volume} {139}},\ \bibinfo {pages}
  {A1844} (\bibinfo {year} {1965})}\BibitemShut {NoStop}%
\bibitem [{\citenamefont {Uba}\ \emph {et~al.}(2017)\citenamefont {Uba},
  \citenamefont {Uba},\ and\ \citenamefont {Antonov}}]{Uba2017}%
  \BibitemOpen
  \bibfield  {author} {\bibinfo {author} {\bibfnamefont {L.}~\bibnamefont
  {Uba}}, \bibinfo {author} {\bibfnamefont {S.}~\bibnamefont {Uba}},\ and\
  \bibinfo {author} {\bibfnamefont {V.~N.}\ \bibnamefont {Antonov}},\
  }\bibfield  {title} {\bibinfo {title} {Magneto-optical kerr spectroscopy of
  noble metals},\ }\href {https://doi.org/10.1103/PhysRevB.96.235132}
  {\bibfield  {journal} {\bibinfo  {journal} {Phys. Rev. B}\ }\textbf {\bibinfo
  {volume} {96}},\ \bibinfo {pages} {235132} (\bibinfo {year}
  {2017})}\BibitemShut {NoStop}%
\bibitem [{\citenamefont {Yu}\ \emph {et~al.}(2021{\natexlab{b}})\citenamefont
  {Yu}, \citenamefont {Wang}, \citenamefont {Zhang}, \citenamefont {Sander},
  \citenamefont {Ni}, \citenamefont {Lu}, \citenamefont {Ma}, \citenamefont
  {Wang}, \citenamefont {Zhao}, \citenamefont {Chen}, \citenamefont {Jiang},
  \citenamefont {Zhang}, \citenamefont {Yang}, \citenamefont {Zhou},
  \citenamefont {Dong}, \citenamefont {Johnson}, \citenamefont {Graf},
  \citenamefont {Hu}, \citenamefont {Gao},\ and\ \citenamefont
  {Zhao}}]{Yu2021a}%
  \BibitemOpen
  \bibfield  {author} {\bibinfo {author} {\bibfnamefont {L.}~\bibnamefont
  {Yu}}, \bibinfo {author} {\bibfnamefont {C.}~\bibnamefont {Wang}}, \bibinfo
  {author} {\bibfnamefont {Y.}~\bibnamefont {Zhang}}, \bibinfo {author}
  {\bibfnamefont {M.}~\bibnamefont {Sander}}, \bibinfo {author} {\bibfnamefont
  {S.}~\bibnamefont {Ni}}, \bibinfo {author} {\bibfnamefont {Z.}~\bibnamefont
  {Lu}}, \bibinfo {author} {\bibfnamefont {S.}~\bibnamefont {Ma}}, \bibinfo
  {author} {\bibfnamefont {Z.}~\bibnamefont {Wang}}, \bibinfo {author}
  {\bibfnamefont {Z.}~\bibnamefont {Zhao}}, \bibinfo {author} {\bibfnamefont
  {H.}~\bibnamefont {Chen}}, \bibinfo {author} {\bibfnamefont {K.}~\bibnamefont
  {Jiang}}, \bibinfo {author} {\bibfnamefont {Y.}~\bibnamefont {Zhang}},
  \bibinfo {author} {\bibfnamefont {H.}~\bibnamefont {Yang}}, \bibinfo {author}
  {\bibfnamefont {F.}~\bibnamefont {Zhou}}, \bibinfo {author} {\bibfnamefont
  {X.}~\bibnamefont {Dong}}, \bibinfo {author} {\bibfnamefont {S.~L.}\
  \bibnamefont {Johnson}}, \bibinfo {author} {\bibfnamefont {M.~J.}\
  \bibnamefont {Graf}}, \bibinfo {author} {\bibfnamefont {J.}~\bibnamefont
  {Hu}}, \bibinfo {author} {\bibfnamefont {H.-J.}\ \bibnamefont {Gao}},\ and\
  \bibinfo {author} {\bibfnamefont {Z.}~\bibnamefont {Zhao}},\ }\href@noop {}
  {\bibinfo {title} {Evidence of a hidden flux phase in the topological
  kagom\'e metal {CsV$_3$Sb$_5$}}} (\bibinfo {year} {2021}{\natexlab{b}}),\
  \Eprint {https://arxiv.org/abs/2107.10714} {arXiv:2107.10714
  [cond-mat.supr-con]} \BibitemShut {NoStop}%
\end{thebibliography}%

\onecolumngrid
\newpage

\renewcommand{\thefigure}{S\arabic{figure}}
\renewcommand{\theequation}{S.\arabic{equation}}
\renewcommand{\thetable}{S\arabic{table}}

\renewcommand{\thefootnote}{\fnsymbol{footnote}}

\begin{center}
	\textbf{SUPPLEMENTAL INFORMATION for} \\
	\vspace{1em}
	\textbf{High Resolution Polar Kerr Effect Studies of CsV$_3$Sb$_5$ and ScV$_6$Sn$_6$ Below the Charge Order Transition}\\
	
	\fontsize{9}{12}\selectfont
	
	\vspace{2em}
	David R. Saykin,$^{1,2,3}$ Qianni Jiang,$^{3,4}$ Zhaoyu Liu,$^{5}$ Chandra Shekhar,$^{6}$ Claudia Felser,$^{6}$ Jiun-Haw Chu,$^{5}$ and Aharon Kapitulnik,$^{1,2,3,4}$\\
	\vspace{1em}
	$^1${\it Geballe Laboratory for Advanced Materials, Stanford University, Stanford, CA 94305, USA}\\
	$^2${\it Department of Physics, Stanford University, Stanford, CA 94305, USA}\\
	$^3${\it Stanford Institute for Materials and Energy Sciences, SLAC National Accelerator Laboratory, 2575 Sand Hill Road, Menlo Park, CA 94025, USA.}
	$^4${\it Department of Applied Physics, Stanford University, Stanford, CA 94305, USA}\\
	$^5${\it Department of Physics, University of Washington, Seattle, Washington, 98195, USA}\\
	$^6${\it Max Planck Institute for Chemical Physics of Solids, 01187 Dresden, Germany}\\
\end{center}

\section*{List of supplemental content:}
\begin{itemize}
	\item [S1.] Samples
	\item [S2.] Zero-area-loop Fiber-optic Sagnac Interferometer (ZALSI)
	\item [S3.] Detection of CDW with coherent reflection ratio $P_2/P_0$
	\item [S4.] General Considerations for MO effects
	\item [S5.] Resistivity Measurements
\end{itemize}
\bigskip

\setlength{\baselineskip}{1.1\baselineskip}  
\setcounter{section}{0}
\renewcommand{\thesection}{S\arabic{section}}

\section{Samples}\label{app:samples}
\setcounter{equation}{0}\renewcommand{\theequation}{A.\arabic{equation}}
In the present study \CVS samples were grown in two different laboratories first set of samples used for the  Sagnac measurements were grown in Dresden, while second set of samples, primarily used for strain-effect measurements were grown at the University of Washington. Both set of crystals showed similar characteristics in structure, transport and optical response measurements.

\section{Zero-area-loop Fiber-optic Sagnac Interferometer (ZALSI)}\label{app:ZALSI}
\setcounter{equation}{0}\renewcommand{\theequation}{A.\arabic{equation}}

Samples were measured using zero loop area fiber Sagnac interferometers (ZALSI) \cite{Xia2006} using $30$ $\mu$W optical power at 1550 nm wavelength with phase modulation at $\omega=5$ MHz. Two low-coherence light waves of right and left circularly polarizations were sent to the sample. And the non-reciprocal phase difference $\varphi_{nr} = 2 \theta_K$ between the two lights acquired upon reflection was detected with lock-in amplifiers. By construction, unlike a standard ellipsometer, this approach fundamentally rejects polarization rotations due to non-TRSB effects such as linear and circular birefringence and dichroism that could mimic a TRSB Kerr signal. In addition, by reducing the Sagnac loop to zero area within a single fiber, it also rejects a background Sagnac signal from earth rotation, which breaks time-reversal symmetry and is the basis for fiber gyroscopes. 

\begin{figure}[h]
	\includegraphics[width=0.8\columnwidth]{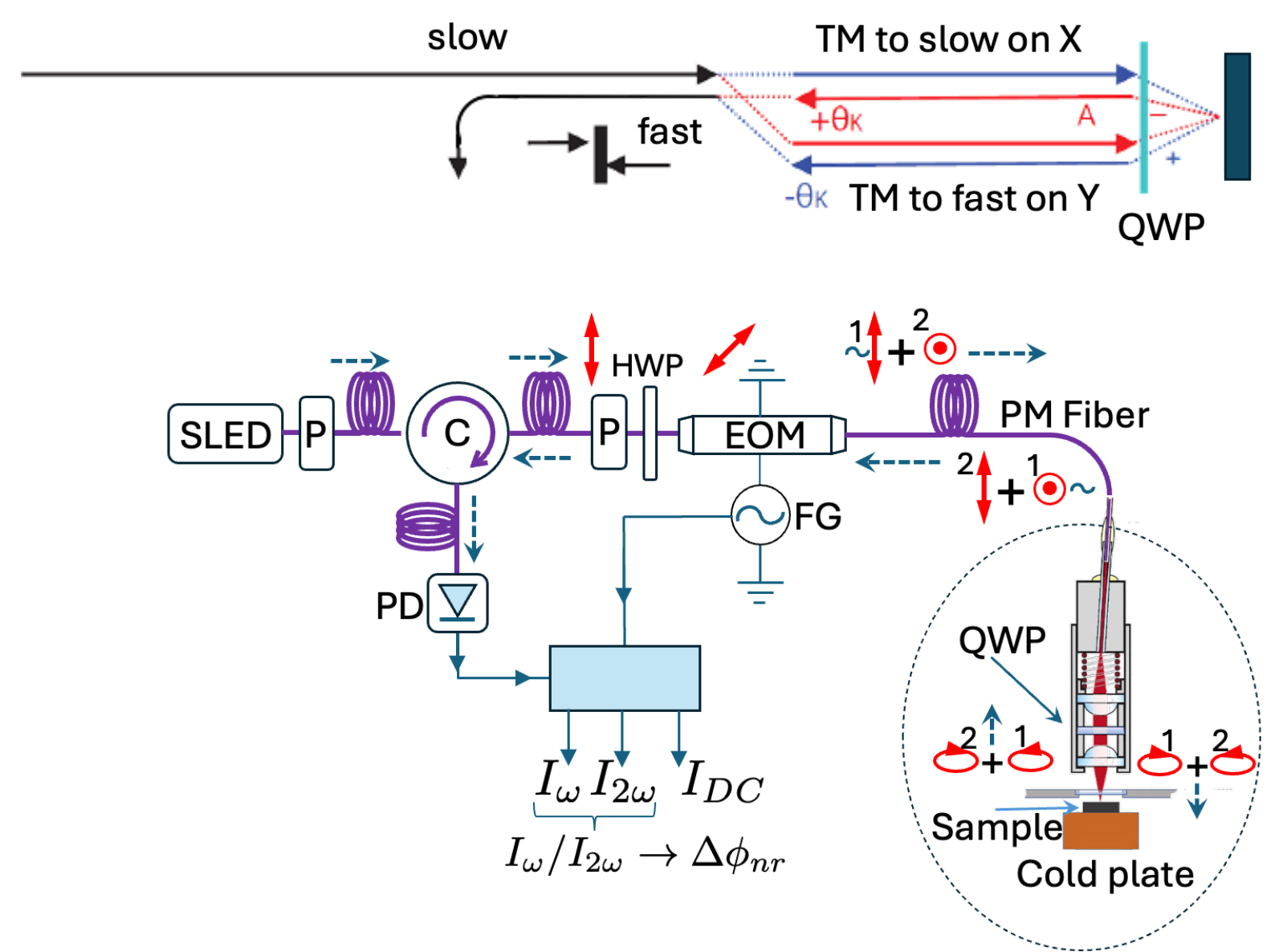}
	\caption{Schematics of the zero-area-loop fiber-optic Sagnac interferometer (ZALSI). Top figure is a schematic of the round trip of the two counter-propagating beams.}
\label{fig:ZALSI}
\end{figure}

A schematics of ZALSI as well as polarization states at each point are shown in Fig.~\ref{fig:ZALSI}. The beam of light polarized at 45$\rm ^o$  to the axis of a electro-optic modulator (EOM), which generates $5$ M\rm Hz time-varying differential phase shifts $\phi_m$ along its two major axis and split the light into two beams of roughly equal powers. The two beams are then launched into the fast and slow axes respectively of a polarization maintaining (PM) fiber. Upon exiting the fiber, the two orthogonally polarized beams are converted into right- and left-circularly polarized light by a quarter-wave ($\lambda /4$) plate, and are then focused onto the sample. The non-reciprocal phase shift $\phi_{nr}$ between the two circularly polarized beams upon reflection from the sample is twice the Kerr rotation ($\varphi_{nr} = 2 \theta_K$). The same quarter-wave plate converts the reflected beams back into linear polarization, but with a net 90$^o$ rotation of the polarization axis. The two beams then travel back through the PM fiber and the EOM with exchanged axes before they arrive again at the polarizer. At this point, the two beams have gone through exactly the same path but in opposite directions, except for a phase difference of  $\phi_{nr} = 2 \theta_K$ from reflection off of the sample. The two beams are once again coherent, and interfere to produce an elliptically polarized beam, whose in-plane component is routed by the circulator to the photodetector. Lock-in detection was used to measure the average (DC) power ($P_0$), the first harmonics ($P_1$), and the second harmonics ($P_2$) of the detected optical power $P(t)$:

\begin{equation}
\centering
P(t)=\frac{1}{2}P[1+\cos(\varphi_{nr}+2\phi_m\sin(\omega t))]
\end{equation}

where $P$ is the returned power without modulation, and depends on focus of the objective lens and sample reflectivity. $P(t)$ can be further expanded into Fourier series of $\omega$ if we keep $\varphi_{nr}$ as a slowly time-varying quantity compared to $\omega$:

\begin{eqnarray}
\centering
P(t)/P &=& [1+J_0(2\phi_m)]/2 \nonumber\\
&&+ (\sin (\varphi_{nr}) J_1(2\phi_m))\sin(\omega t) \nonumber\\ 
&&+ (\cos (\varphi_{nr}) J_2(2\phi_m))\cos(2\omega t)\nonumber\\
&&+ 2J_3(2\phi_m)\sin(3\omega t)\nonumber\\
&&+ ... \label{eq:fourier}
\end{eqnarray}

where $J_1$ and $J_2$ are Bessel functions. Therefore, the detected powers $P_0$, $P_1$ and $P_2$ are: 

\begin{equation}
\centering
P_0/P=[1+J_0(2\phi_m)]/2
\end{equation}

\begin{equation}
\centering
P_1/P=(\sin (\varphi_{nr}) J_1(2\phi_m))
\end{equation}

\begin{equation}
\centering
P_2/P=(\cos (\varphi_{nr}) J_2(2\phi_m))
\end{equation}

Hence Kerr signal $\theta_K=\varphi_{nr}/2 $ can be obtained using equation \ref {eq:Sagnac_kerr}, which is independent of optical power, sample reflectivity and focus of the objective lens. For optimal $\theta_K$ sensitivity $\phi_m$ is often chosen to be close to 0.92. 

\begin{equation}
\theta_K = \frac{1}{2} \tan^{-1}\left[ \frac{J_2(2\phi_m)P_1}{J_1(2\phi_m) P_2}\right]
\label{eq:Sagnac_kerr}
\end{equation}

\section{Detection of CDW with coherent reflection ratio $P_2/P_0$}\label{app:jones}
\setcounter{equation}{0}\renewcommand{\theequation}{B.\arabic{equation}}

In addition to the Kerr signal, we record the total (P0) and the coherence (P2) parts of the reflected optical power, as their ratio serves as a measure of the linear and/or circular birefringence. The above calculations of the ZALSI assume perfect retardance of the quarter-wave plate and absence of either linear or circular birefringence and dichroism of the sample. In reality, commercial zero-order quarter-wave plates have a typical retardance error of 1$\%$ even for normal incidence, and samples such as \CVS display birefringence and/or dichroism. As a result, the reflected beams, after passing the quarter-wave plate again, become elliptical instead of being perfectly linearly polarized. And a small fraction of the light will be incoherent with the major beams and thus won't participate in the interference. These incoherent components will not be captured by $P_1$ or $P_2$, but will still be detected as part of the average power $P_0$. And pre-factors need to be added to the formulas for $P_0$, $P_1$ and $P_2$:

\begin{equation}
\centering
P_0/P= (1+A_0) [1+J_0(2\phi_m)]/2
\end{equation}

\begin{equation}
\centering
P_1/P= (1+A_1) (\sin (\varphi_{nr}) J_1(2\phi_m))
\end{equation}

\begin{equation}
\centering
P_2/P=(1+A_2) (\cos (\varphi_{nr}) J_2(2\phi_m))
\end{equation}

where $A_0$, $A_1$ and $A_2$ are small correction pre-factors for sample birefringence and/or dichroism, and retardance error of the wave plate. The Kerr signal $\theta_K$ can be obtained using updated equation \ref {eq:Sagnac_kerr1}, with a small correction to the scaling factor. There is no change to the zero point of $\theta_K$, which is guaranteed by the symmetry of the interferometer. 

\begin{equation}
\theta_K = \frac{1}{2} \tan^{-1}\left[ \frac{(1+A_2)J_2(2\phi_m)P_1}{(1+A_1)J_1(2\phi_m) P_2}\right]
\label{eq:Sagnac_kerr1}
\end{equation}

On the other hand, a change in sample birefringence and/or dichroism will induce changes to $P_0$ and $P_2$. However, as previously mentioned, they are also dependent on $P$, which changes with focus of the objective lens and sample reflectivity. 

\begin{equation}
P_2/P_0 = \frac{(1+A_2)J_2(2\phi_m)}{(1+A_0)(1+J_0(2\phi_m))}
\label{eq:ratio}
\end{equation}

Their ratio $P_2/P_0$ is independent of these factors and represents the ratio between the coherent and the total optical powers, dubbed "coherent reflection ratio". Since the wave plate retardance error is a slow varying quantity usually dominated the slow drifts of its tilt and rotation, $P_2/P_0$ can be used to measure that change of sample birefringence and/or dichroism during temperature sweeps.

\section{General Considerations for MO effects}\label{app:MO}
In general Magneto-optical (MO) effects appear because in the presence of magnetism right and left circularly polarized lights propagate differently in solids. When a magnetic field is applied to a diamagnetic insulating solid, magneto-optical effects will originate from the direct effect of the magnetic field on the orbital electronic motion. On the other hand, for ferromagnetic materials, or paramagnetic materials at low temperatures (when their Curie susceptibility is large enough), the effect of the magnetic field on the orbital motion is negligible compared with effects associated with spin-orbit interaction  \cite{Pershan1967}. For simple metals, far from plasma frequency resonances we expect that the main contribution to Kerr response is dominated by off-diagonal intraband Drude-type transitions  (i.e. originating from optical conductivity terms $\sigma_{xy}(\omega)=\sigma_0(\omega_c\tau)/[(1-i\omega\tau)^2+(\omega_c\tau)^2]$, where $\omega_c=eH/m^*c$ is the cyclotron frequency,   $ \sigma_0$ is the DC Drude conductivity and $\tau$ is the scattering time). For example, in  Al and Ag \cite{Stern1964} and nobel metals including Cu and Au \cite{McGroddy1965} these effects were measured and recently calculated, showing that for energies below $\sim 1.5$ eV the Kerr rotation is of order $\sim 10^{-9}$ rad/Oe \cite{Uba2017}. In the absence of magnetic polarization, the orbital and spin Zeeman terms will contribute off-diagonal terms through interband transitions, which for the above simple metals are at least an order of magnitude smaller. Taking into account optical and transport measurements on \CVS, both effects are expected to yield an even smaller response, which will not be detectable for the magnetic fields we used with the ZALSI experiments.

\section{Resistivity Measurements}\label{app:R}

\begin{figure}[h]
	\includegraphics[width=0.8\columnwidth]{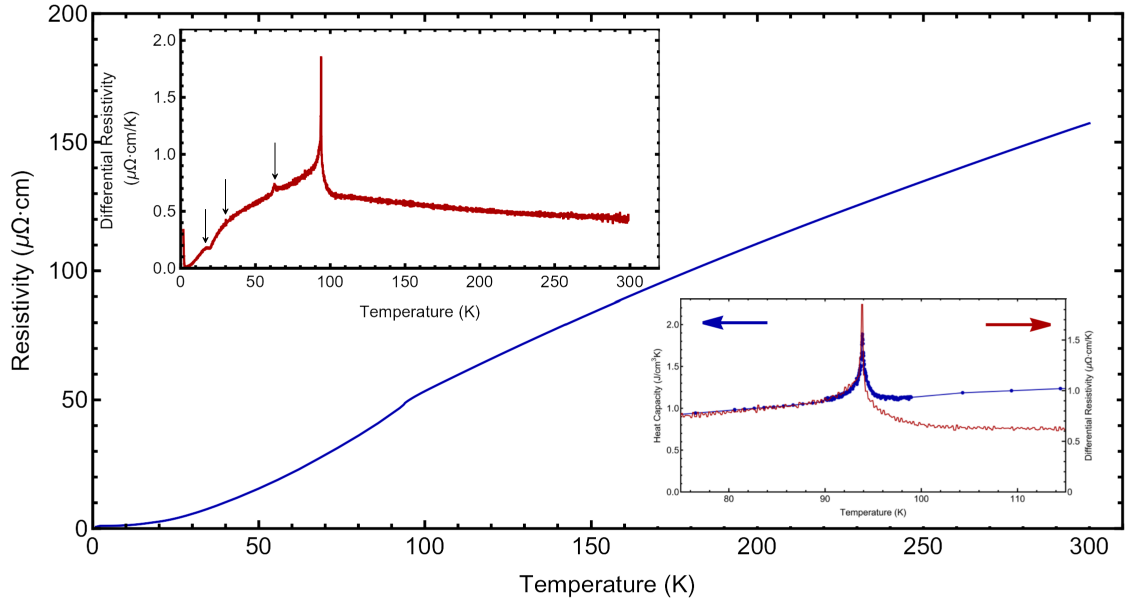}
	\caption{Typical resistivity vs. temperature plot for our samples. Top inset shows the derivative of the resistivity with pronounced anomalies at $\sim 65$ K and $\sim 18$ K, and a weaker anomaly at $\sim 30$ K (see arrows). Bottom inset shows the resistivity anomaly laid on the specific heat anomaly, showing similar behavior below $T_{CDW}$.}
\label{resistivityCVS}
\end{figure}

DC resistance was measured on typical \CVS samples from the same batch as the Kerr-measurements samples and shown in Fig.~\ref{resistivityCVS}. A kink at $T_{CDW}$ is clearly visible in the resistance, marking the charge density transition. The exact value of $T_CDW$ is determined by the temperatures of the peaks in the first derivative $dR/dT$ curve, and agree with literature data \cite{Ortiz2020} within $1$ K.  In the $dR/dT$ curve there are  additional features of dispersive line shape centered at $T_A\approx 65$ K and $T_B\approx 18$ K. We note that different measurements on different crystals often observed subtle structural effects at $\sim 70$ K \cite{Yu2021a}, or $\sim 35$ K \cite{Guo2022}. 


\end{document}